\documentclass[10pt,a4paper]{article}
\usepackage{float}
\usepackage[margin=2.0cm]{geometry}
\usepackage[T1]{fontenc}
\usepackage{lmodern}
\usepackage{amsmath,amssymb,bm}
\usepackage{booktabs}
\usepackage{array}
\usepackage{longtable}
\usepackage{multirow}
\usepackage{enumitem}
\usepackage{microtype}
\usepackage{xcolor}
\usepackage{graphicx}
\usepackage{subcaption}
\usepackage{cite}
\usepackage[
    colorlinks=true,
    linkcolor=blue,
    citecolor=blue,
    urlcolor=blue
]{hyperref}
\setlist[itemize]{topsep=0.25em,itemsep=0.2em,leftmargin=1.6em}
\setlist[enumerate]{topsep=0.25em,itemsep=0.25em,leftmargin=1.8em}

\usepackage{orcidlink}

\title{\textbf{Dynamical stability and observational constraints on a
late-time dark-matter particle-creation model}}

\author{
Anirban Chatterjee\,\orcidlink{0000-0002-8904-7133}
\thanks{ 
\texttt{iitkanirbanc@gmail.com};
\texttt{anirban.chatterjee@gm.rkmvu.ac.in}
}
\\[0.5em]
\small Department of Physics, School of Mathematical Sciences\\
\small Ramakrishna Mission Vivekananda Educational and Research Institute\\
\small Belur Math, Howrah 711202, West Bengal, India
}

\date{}

\begin{document}

\maketitle

\begin{abstract}
We investigate the cosmological dynamics of gravitational
particle creation in a spatially flat FLRW Universe comprising
radiation, baryons, dark matter (DM), and a separately conserved
dark-energy component. The DM sector is described as an effective
open thermodynamic fluid incorporating intrinsic pressure and
particle-creation-induced negative pressure. Introducing the
compact variable $u=H_0/(H_0+H)$, we construct a four-dimensional
autonomous system and systematically investigate its critical
points, dynamical stability, and physical admissibility.
The model is constrained using three observational combinations:
PPS+BAO+OHD, Union3+BAO+OHD, and DES-DOV+BAO+OHD.
The observationally preferred parameter regions exclude the
nontrivial DM-dominated de Sitter-like branches from the physical
phase space, while identifying a stable, physically admissible,
dark-energy-dominated accelerating attractor. The reconstructed
cosmological evolution consistently reproduces the
radiation--matter--acceleration sequence, with a transition
redshift $z_{\rm tr}\simeq0.71$--$0.73$.
Posterior-level analysis reveals that approximately
$92\%$--$97\%$ of the samples support this accelerating attractor.
The asymptotic evolution approaches a quintessence-like
power-law expansion with $H\rightarrow0$, rather than a
de Sitter state. These findings establish a connection between
observational constraints, gravitational particle creation,
and the global dynamical structure governing the distant
cosmic future.
\end{abstract}

\section{Introduction}
\label{sec:introduction}

The discovery of the late-time accelerated expansion of the Universe from Type Ia supernova observations \cite{ref:Riess98,ref:Perlmutter} established one of the major open problems of modern cosmology. Within the standard $\Lambda$CDM framework, this acceleration is attributed to a cosmological constant, while the clustering matter sector is dominated by pressureless cold dark matter. The model provides an excellent description of a wide range of cosmological observations, including the cosmic microwave background, baryon acoustic oscillations, and large-scale structure \cite{ht1,SDSS:2005xqv,DESI:2024mwx}. Nevertheless, the physical nature of both dark energy and dark matter remains unknown, and the extremely small observed value of the vacuum energy compared with natural theoretical expectations continues to motivate alternatives to a strictly constant vacuum component. A broad range of possibilities has therefore been investigated, including dynamical dark energy \cite{Zlatev:1998tr,Copeland:1997et,Peccei:1987mm,Ford:1987de,ArmendarizPicon:1999rj,ArmendarizPicon:2000ah}, interacting dark sectors \cite{Pourtsidou:2016ico,Bandyopadhyay:2017igc,Bandyopadhyay:2018zlz,Bandyopadhyay:2019vdd,Chatterjee:2022uyw,Yang:2018uae}, phenomenological varying EoS parametrization models \cite{Bandyopadhyay:2019ukl,Chatterjee:2026sde} and modified-gravity models \cite{fr1,fr2,fr3,fr4,fr5,fr6,fr7}. Another possibility, conceptually distinct from introducing a new interaction between dark matter and dark energy, is that the matter sector itself behaves as an open thermodynamic system in which gravitationally induced particle creation modifies the effective cosmic pressure.

Particle creation in an expanding spacetime has a long history, beginning with the quantum-field-theoretic description of particles generated by a time-dependent gravitational background \cite{Parker1968,KolbLong2024}. At the macroscopic level, cosmological particle production can be described within non-equilibrium thermodynamics by allowing the particle current to be non-conserved \cite{Prigogine1988,Calvao1992,Lima1996,Zimdahl2001}. For adiabatic particle creation, in which the entropy per particle remains constant while the total entropy increases, the source term in the particle balance equation can be represented by an effective negative creation pressure. The matter component can therefore acquire a negative effective equation of state without requiring a fundamental negative-pressure matter field. This mechanism has been extensively explored in cosmology and can, for suitable creation rates, contribute to or even generate late-time accelerated expansion \cite{Steigman2009,Lima2010,Basilakos2010,Jesus2011,Ramos2014}. In this sense, gravitational particle creation provides a natural framework in which non-equilibrium matter dynamics can influence the background expansion.

Most particle-creation cosmologies depend crucially on the phenomenological form adopted for the creation rate $\Gamma$ \cite{Steigman2009,Lima2010,Basilakos2010,Jesus2011,Ramos2014,LimaLima2026}. From the viewpoint of late-time cosmology, a desirable parametrization should satisfy two basic requirements. First, particle creation should be sufficiently suppressed in the high-redshift regime so that the conventional radiation- and matter-dominated eras are recovered. Second, it should be allowed to become dynamically relevant as the Hubble rate decreases, thereby affecting the recent evolution without significantly modifying the well-established early-time sequence. Motivated by these considerations, we formulate the dark-matter sector in terms of the compact variable $u=\frac{H_0}{H_0+H}$, which maps the expanding background evolution onto the finite interval $0\leq u\leq1$. The limit $u\rightarrow0$ corresponds to the high-$H$ regime, the present epoch is fixed by $u_0=1/2$, and $u\rightarrow1$ corresponds to the compact future boundary $H\rightarrow0$. Besides being convenient for the dynamical-system formulation, this variable makes the asymptotic behavior of the dark-matter sector transparent.

Within this compact description, we consider a minimal late-time realization in which the intrinsic dark-matter equation of state is linear in $u$, while the normalized particle-creation strength is quadratic. The effective dark-matter equation of state then takes the form $\omega_{m,\rm eff}(u)=2\omega_{m0}u-4\gamma_0u^2$.  This choice ensures that both the intrinsic pressure and the creation contribution vanish in the high-$H$ limit, while the latter is suppressed more rapidly because of its quadratic dependence on $u$. Consequently, the dark-matter sector continuously approaches a pressureless and non-creating regime toward high redshift, whereas deviations from this limit are naturally confined to late times. At the present epoch the two contributions combine according to
$\omega_{m,\rm eff}^{0}=\omega_{m0}-\gamma_0$, so the similar present-day effective dark-matter behavior can arise from different combinations of intrinsic pressure and creation amplitude. This relation later appears directly in the posterior correlations.

The dark-matter sector is motivated by an effective covariant open-fluid construction based on the variational formulation of relativistic fluids \cite{Schutz1970,Brown1993}. In this description, the non-conservation of the particle current is generated by an effective source term, while the number-density dependence of the energy density determines the intrinsic pressure. Since the compact variable used here depends on the expansion scalar, the construction is intended as an effective covariant realization of the homogeneous background model rather than as a unique microscopic completion. The analysis is therefore deliberately focused on background dynamics, physical phase-space structure, and late-time observational constraints.

A central objective of the present work is to connect this open-fluid description with a complete dynamical-system analysis. Dynamical-systems techniques are particularly useful in cosmology because they permit the global expansion history to be organized in terms of critical points, invariant sets, and their stability properties \cite{Copeland:1997et,Bahamonde2018,Fang:2014qga,
Hussain:2023kwk, Bhattacharya:2022wzu, Chatterjee:2021ijw,Chatterjee:2021hhj,Chatterjee:2024duy,Chatterjee:2025htd}. We rewrite the background equations as a four-dimensional autonomous system in
$(\Omega_{\rm DM},\Omega_r,\Omega_b,u)$, with the dark-energy fraction determined by the Friedmann constraint. Retaining the baryonic variable explicitly is important because in the high-$H$ limit the effective dark-matter pressure vanishes and the baryonic and dark-matter components become dynamically degenerate. This gives rise to a continuous matter critical line and requires the full four-dimensional Jacobian for a consistent stability classification.

The resulting phase space contains the standard radiation, matter, and dark-energy-dominated configurations as well as nontrivial dark-matter-dominated branches associated with the creation sector. Particular attention is paid to the points $P_7$ and $P_8$, which formally exhibit de Sitter-like effective behavior, and to the dark-energy-dominated points $P_9$ and $P_{10}$. A key aspect of the analysis is that mathematical stability is not by itself sufficient to establish cosmological viability. A critical point may have a stable eigenvalue spectrum while lying outside the physical compact interval $0\leq u\leq1$, and an accelerating point may still be a saddle. The simultaneous requirements of acceleration, physical admissibility, and linear stability therefore provide a more stringent criterion for identifying the actual late-time endpoint.

The theoretical model is confronted with three observational combinations of late-time background observations. We consider the Pantheon+ supernova compilation \cite{Scolnic:2021amr,Brout2022}, Union3 \cite{Rubin:2023jdq}, and the DES five-year supernova sample \cite{DES:2024jxu,DES:2025sig}, each combined with baryon acoustic oscillation measurements and observational Hubble data. The BAO sector is based on the first-year DESI measurements \cite{DESI:2024mwx,DESI:2024jxi,Lu:2024hvv,Colgain:2024xqj}, while direct measurements of $H(z)$ are supplied by the cosmic-chronometer technique \cite{Moresco2016,Jimenez2003,Simon2005,Stern2010,Moresco2012,Moresco2015,Ratsimbazafy2017,Borghi2022}. The analysis is intentionally restricted to late-time observables. In particular, the BAO scale is represented through the phenomenological combination $hr_d$, avoiding the imposition of a specific early-Universe calibration of the sound horizon. This allows the observational constraints to remain closely tied to the late-time phenomenology of the particle-creation model.

The final posterior chains are expressed in the common parameter basis
$(\Omega_{m0},h,hr_d,\omega_{\rm de},\omega_{m,\rm eff}^{0},\gamma_0)$.
Since the autonomous system is naturally formulated in terms of the intrinsic present-day dark-matter parameter, this quantity is reconstructed sample by sample through $
\omega_{m0}
=
\omega_{m,\rm eff}^{0}+\gamma_0$. 
The sample-by-sample transformation is essential because $\omega_{m,\rm eff}^{0}$ and $\gamma_0$ are strongly anticorrelated. The three observational combinations yield broadly overlapping posterior regions, with the most visible differences appearing in $h$ and $\omega_{\rm de}$. In all cases the preferred present-day effective dark-matter equation-of-state parameter is mildly negative, whereas the reconstructed intrinsic dark-matter parameter remains mildly positive. The posterior for the creation amplitude is shifted toward positive values, although its broad posterior implies that the present background data should not be interpreted as establishing a high-significance detection of particle creation.

The observational inference is then propagated back into the phase-space structure. This allows us to address two complementary questions: which critical point is selected by the exact best-fit cosmology, and how robust that classification remains across the full posterior volume. At the three exact best-fit parameter sets, the nontrivial point $P_7$ lies outside the physical domain at $u<0$, while $P_8$ lies beyond the upper compact boundary at $u>1$. The formally accelerating point $P_6$ remains a saddle, and the dark-energy point $P_9$ is likewise saddle in the relevant direction. In contrast, $P_{10}=(0,0,0,1)$ satisfies both the acceleration and stability requirements for all three observational calibrations. Since $u=1$ corresponds to $H\rightarrow0$, this state is not a finite-$H$ de Sitter solution but a compactified, dark-energy-dominated, quintessence-like accelerating endpoint.

The posterior-level dynamical analysis further shows that this conclusion is not confined to isolated maximum-likelihood points. Most of the retained samples from all three datasets lie in the stable-and-accelerating region associated with $P_{10}$, with posterior support exceeding $90\%$ in every case. The reconstructed background evolution is correspondingly similar across the three observational combinations: the expansion histories nearly overlap, and the transition from deceleration to acceleration occurs consistently around $z_{\rm tr}\simeq0.71$--$0.73$. The intrinsic and effective dark-matter equations of state both approach the pressureless limit toward high redshift, while the particle-creation contribution becomes increasingly important only at late times.

The main purpose of this work is therefore to establish a direct connection between three levels of description: the effective open-fluid formulation of dark matter, the global autonomous dynamics, and the observationally preferred parameter region. By combining these ingredients, we show that the mathematically available late-time solutions are strongly restricted once phase-space physicality and observational calibration are imposed simultaneously. The analysis selects the dark-energy-dominated point $P_{10}$ as the physically relevant future attractor, while the formal dark-matter-dominated de Sitter-like branches are excluded at the observational best fits. This provides a unified dynamical and observational characterization of late-time dark-matter particle creation.
 
The remainder of this paper is organized as follows. Section~\ref{sec:theory} introduces the effective open-fluid formulation of the dark-matter sector, the compact variable $u$, and the corresponding particle-creation parametrization. Section~\ref{sec:dynamical_system} develops the four-dimensional autonomous system, derives the critical points and Jacobian eigenvalues, and discusses their physical interpretation and stability. Section~\ref{sec:observations} presents the late-time observational datasets, the cosmological parameterization, and the statistical methodology used for parameter inference. The observational constraints are then implemented in the dynamical framework in Sec.~\ref{sec:obs_dynamics}, where we analyze the physicality and stability of the late-time critical points, reconstruct the background evolution, quantify the posterior support for the accelerating attractor, and study the global phase-space and density evolution. Finally, Sec.~\ref{sec:conclusions} summarizes the main results and outlines possible extensions of the model.

\section{Theoretical Framework}
\label{sec:theory}

We consider a spatially flat Friedmann--Lemaître--Robertson--Walker
(FLRW) spacetime,
\begin{equation}
ds^2=-dt^2+a^2(t)d{\bf x}^2,
\qquad
H=\frac{\dot a}{a},
\end{equation}
comprising radiation, baryons, dark matter (DM), and a dark-energy
component characterized by a constant equation-of-state parameter
$\omega_{\rm de}$. Radiation, baryons, and dark energy are assumed
to be separately conserved, while the DM sector is treated as an
effective open thermodynamic fluid undergoing gravitational particle
creation \cite{Prigogine1988,Calvao1992,Lima1996,Zimdahl2001,
Steigman2009,Lima2010,Ramos2014}.

The individual energy conservation equations are given by
\begin{align}
\dot{\rho}_r+4H\rho_r &= 0, \qquad
\dot{\rho}_b+3H\rho_b = 0,\\
\dot{\rho}_{\rm de}+3H(1+\omega_{\rm de})\rho_{\rm de} &= 0, \qquad
\dot{\rho}_m+3H(\rho_m+p_m+p_c) = 0.
\end{align} 
where $\rho_r$, $\rho_b$, $\rho_{\rm de}$, and $\rho_m$ denote the
energy densities of radiation, baryons, dark energy, and dark
matter, respectively. Here, $p_m=\omega_m\rho_m$ represents the
intrinsic DM pressure, while $p_c$ denotes the effective creation
pressure associated with gravitational particle production. Inclusion of a separately conserved dark-energy component provides a broader dynamical framework for investigating late-time cosmic acceleration in the presence of gravitational particle creation. Furthermore, it allows us to explore the
asymptotic cosmic evolution and long-term fate of the Universe.

This framework modifies the DM energy balance through an effective
negative creation pressure without introducing any direct
non-gravitational interaction between DM and dark energy.
Consequently, gravitational particle creation provides a
phenomenological mechanism through which non-equilibrium
dark-matter dynamics can influence the late-time cosmological
expansion.
The DM particle current is defined as $N^\mu=nU_m^\mu$, where
$n$ denotes the particle-number density and $U_m^\mu$ is the
DM four-velocity. Its non-conservation is described by
\begin{equation}
\nabla_\mu N^\mu=n\Gamma_N,
\qquad
\dot n+3Hn=\Gamma_N n,
\end{equation}
where $\Gamma_N$ represents the particle-number creation rate.
For adiabatic particle creation, the entropy per particle remains
constant, although the total entropy may increase through particle
production. The Gibbs relation then gives
\begin{equation}
\dot\rho_m+3H(1+\omega_m)\rho_m
=\Gamma\rho_m,
\qquad
\Gamma=(1+\omega_m)\Gamma_N,
\label{eq:dm_balance}
\end{equation}
where $\omega_m=p_m/\rho_m$ denotes the intrinsic DM
equation-of-state parameter. The corresponding creation pressure
and effective equation of state are
\begin{equation}
p_c=-\frac{\Gamma}{3H}\rho_m,
\qquad
\omega_{m,\rm eff}
=\omega_m-\frac{\Gamma}{3H}.
\label{eq:wmeff_general}
\end{equation}
Thus, the effective DM pressure contains contributions from both
the intrinsic thermodynamic properties of the fluid and the
non-equilibrium particle-production process.

At the covariant level, the non-conservation of the particle
current can be motivated within effective relativistic fluid
formulations, including the Schutz--Sorkin approach
\cite{Schutz1970,Brown1993}. In the present framework, however,
the particle-production source and its associated creation
pressure are introduced phenomenologically through the
open-system balance equations, without assuming a unique
microscopic action. The model is therefore regarded as an
effective covariant open-fluid description, with the present
investigation restricted to homogeneous cosmological dynamics.

The choice of the phenomenological parametrization is guided
by the requirement that the conventional radiation- and
matter-dominated epochs remain approximately unaffected,
while deviations from pressureless dark-matter behavior may
become relevant during the recent cosmological evolution.
In particular, the intrinsic DM pressure and normalized
particle-creation strength should both be negligible in
the early high-$H$ regime, without imposing their complete
absence at late times. A convenient realization of these
requirements is obtained by introducing a bounded variable
that continuously tracks the expansion history.

Motivated by this consideration, we define the DM expansion
scalar $\Theta_m=\nabla_\mu U_m^\mu$ and introduce
\begin{equation}
u\equiv\frac{H_0}{H_0+\Theta_m/3}.
\end{equation}
On the homogeneous FLRW background, $\Theta_m=3H$, yielding
\begin{equation}
u=\frac{H_0}{H_0+H},
\qquad
\frac{H}{H_0}=\frac{1-u}{u}.
\label{eq:u_definition}
\end{equation}
For an expanding Universe, $0<u<1$, with $u\rightarrow0$
corresponding to the high-$H$ regime and $u\rightarrow1$
to the compactified limit $H\rightarrow0$. At the present
epoch, $u_0=1/2$. This compactification maps the expanding
background history onto a finite interval and provides a
natural basis for parametrizing the dark-matter sector.
Furthermore, polynomial functions of $u$ remain regular
throughout the physical domain, facilitating both the
background reconstruction and subsequent dynamical-system
analysis. To recover pressureless DM in the high-$H$ limit, we impose
$\omega_m(0)=0$. Assuming that the intrinsic equation of
state admits a regular expansion around $u=0$, one may write
\begin{equation}
\omega_m(u)=c_1u+\mathcal O(u^2).
\end{equation}
Retaining the leading non-vanishing contribution and
normalizing the equation of state at the present epoch
through $\omega_m(u_0)=\omega_{m0}$ gives
\begin{equation}
\omega_m(u)=2\omega_{m0}u.
\label{eq:wm_u}
\end{equation}
The linear dependence therefore represents the lowest-order
nontrivial correction to pressureless DM that is regular
throughout the compact interval. It introduces only one
additional intrinsic-pressure parameter while ensuring that
the deviation from pressureless behavior decreases
continuously toward the early Universe. For the particle-creation sector, we similarly require
$\Gamma/(3H)\rightarrow0$ as $u\rightarrow0$. Since the
creation-induced pressure is intended to represent a
late-time modification of the DM dynamics, we further
assume that its normalized contribution is suppressed
more rapidly than the leading intrinsic-pressure correction
in the high-$H$ regime. The simplest polynomial realization
satisfying this requirement is 
\begin{equation}
\frac{\Gamma}{3H}=4\gamma_0u^2
\label{eq:Gamma_u}
\end{equation}
  where $\gamma_0$ denotes the present-day normalized
creation strength,
$\left.\Gamma/(3H)\right|_{u_0}=\gamma_0$.
The quadratic dependence is thus a phenomenological
minimality assumption, rather than a consequence of
an underlying symmetry or microscopic particle-production
theory. In contrast to a constant normalized creation
strength, this parametrization suppresses the effective
creation pressure in the early high-$H$ regime while
allowing it to become dynamically relevant as the
expansion rate decreases. Using Eq.~\eqref{eq:u_definition}, the corresponding
background creation rates become
\begin{equation}
\Gamma(u)=12\gamma_0H_0u(1-u),
\qquad
\Gamma_N(u)=
\frac{12\gamma_0H_0u(1-u)}
{1+2\omega_{m0}u}.
\end{equation}
The associated phenomenological local-expansion
prescription can equivalently be expressed as
$\Gamma=4\gamma_0\Theta_m u^2$.
For $\gamma_0>0$ and $1+\omega_m(u)>0$, the source
describes particle creation, whereas negative
$\gamma_0$ formally corresponds to particle
destruction within the same effective framework.

Combining Eqs.~\eqref{eq:wm_u} and \eqref{eq:Gamma_u},
the effective DM equation-of-state parameter takes
the form
\begin{equation}
\omega_{m,\rm eff}(u)
=2\omega_{m0}u-4\gamma_0u^2,
\label{eq:wmeff_u}
\end{equation}
with the present-day relation
\begin{equation}
\omega_{m,\rm eff}^{0}
=\omega_{m0}-\gamma_0.
\label{eq:wmeff0}
\end{equation}
The first term represents the intrinsic DM pressure,
while the second accounts for the negative creation
pressure. Both contributions vanish as $u\rightarrow0$,
with the quadratic creation term being suppressed at
higher order. Consequently, the model recovers the
pressureless and non-creating DM limit at early times,
while permitting a nontrivial competition between
intrinsic pressure and particle creation during the
late-time evolution. Since the homogeneous expansion
depends on their combined contribution through
$\omega_{m,\rm eff}(u)$, background observations
constrain the effective DM dynamics rather than
independently establishing the underlying
particle-production mechanism.

The background gravitational equations are
\begin{align}
3H^2
&=\kappa^2
\left(\rho_r+\rho_b+\rho_m+\rho_{\rm de}\right),
\label{eq:friedmann}
\\
\dot H
&=-\frac{\kappa^2}{2}
\left[
\frac{4}{3}\rho_r+\rho_b
+(1+\omega_{m,\rm eff})\rho_m
+(1+\omega_{\rm de})\rho_{\rm de}
\right],
\label{eq:Hdot}
\end{align}
where $\kappa^2=8\pi G$. The background model is thus
characterized by the phenomenological parameters
$(\omega_{m0},\gamma_0,\omega_{\rm de})$, together with
the present-day cosmological density parameters.
The adopted linear--quadratic parametrization provides
a minimal extension of pressureless DM, preserving
the conventional early-time behavior while allowing
the effective creation pressure to modify the late-time
expansion history. Although higher-order functions of
$u$ may be considered, they would introduce additional
coefficients and potential parameter degeneracies
without being required by the minimal construction.
The present realization is therefore adopted as a
tractable framework for investigating the global
dynamical structure and observational implications
of late-time gravitational particle creation.

\section{Dynamical-System Analysis}
\label{sec:dynamical_system}

To investigate the global cosmological evolution, we recast the background
equations as a four-dimensional autonomous system, following the standard
dynamical-systems approach widely used in dark-energy and modified-gravity
cosmology
\cite{Copeland:1997et,Bahamonde2018,Fang:2014qga,
Chatterjee:2021ijw,Chatterjee:2021hhj,Chatterjee:2024duy,
Chatterjee:2025htd,fr5,fr6,fr7}.
We introduce the dimensionless variables
\begin{equation}
x\equiv\Omega_{\rm DM}
=\frac{\kappa^2\rho_m}{3H^2},
\qquad
y\equiv\Omega_r
=\frac{\kappa^2\rho_r}{3H^2},
\qquad
b\equiv\Omega_b
=\frac{\kappa^2\rho_b}{3H^2},
\label{eq:dyn_variables}
\end{equation}
together with the compact variable
\begin{equation}
u=\frac{H_0}{H_0+H}.
\end{equation}
The dark-energy density parameter follows from the Friedmann constraint,
\begin{equation}
\Omega_{\rm de}=1-x-y-b,
\end{equation}
while the total matter fraction is
\begin{equation}
\Omega_M=x+b.
\end{equation}
Here $x$ refers exclusively to the DM component, whereas $b$ denotes the
baryonic contribution.

Assuming non-negative energy densities, the compact closure of the physical
phase space is
\begin{equation}
\overline{\Psi}
=
\left\{
(x,y,b,u):
x,y,b\geq0,\;
x+y+b\leq1,\;
0\leq u\leq1
\right\}.
\label{eq:phase_space}
\end{equation}
The boundaries $u=0$ and $u=1$ correspond respectively to the compactified
limits $H\rightarrow\infty$ and $H\rightarrow0$.

Using the effective DM equation-of-state parameter derived in
Sec.~\ref{sec:theory},
\begin{equation}
\omega_{m,\rm eff}(u)
=
2\omega_{m0}u-4\gamma_0u^2,
\end{equation}
the total equation-of-state parameter becomes
\begin{equation}
\omega_{\rm tot}
=
x\left(2\omega_{m0}u-4\gamma_0u^2\right)
+\frac{y}{3}
+\omega_{\rm de}(1-x-y-b).
\label{eq:wtot}
\end{equation}

Introducing the e-folding variable $N=\ln a$, with a prime denoting
$d/dN$, we define
\begin{equation}
\mathcal{E}
\equiv
\frac{H'}{H}
=
\frac{\dot H}{H^2}
=
-\frac{3}{2}(1+\omega_{\rm tot}).
\label{eq:Ecal}
\end{equation}
The deceleration parameter is therefore
\begin{equation}
q
=
-1-\mathcal{E}
=
\frac{1}{2}(1+3\omega_{\rm tot}),
\label{eq:q_parameter}
\end{equation}
so that accelerated expansion occurs whenever
$\omega_{\rm tot}<-1/3$.

Using the conservation equations together with
Eq.~\eqref{eq:Ecal}, the background evolution reduces to
\begin{align}
x'
&=
x\left[
-3(1+2\omega_{m0}u)
+12\gamma_0u^2
-2\mathcal{E}
\right],
\label{eq:xprime}
\\
y'
&=
y(-4-2\mathcal{E}),
\label{eq:yprime}
\\
b'
&=
b(-3-2\mathcal{E}),
\label{eq:bprime}
\\
u'
&=
-\mathcal{E}\,u(1-u).
\label{eq:uprime}
\end{align}
The system is therefore specified by
\begin{equation}
{\bf X}=(x,y,b,u)
\end{equation}
and the parameter set
$(\omega_{m0},\gamma_0,\omega_{\rm de})$.

\subsection{Critical points and linear stability}
\label{subsec:critical_stability}

The critical configurations are obtained by imposing
\begin{equation}
x'=y'=b'=u'=0.
\end{equation}
For an isolated critical point
\begin{equation}
P_i=(x_c,y_c,b_c,u_c),
\end{equation}
we perturb the autonomous variables according to
\begin{equation}
{\bf X}={\bf X}_c+\delta{\bf X},
\end{equation}
which gives, to linear order,
\begin{equation}
\delta{\bf X}'
=
J({\bf X}_c)\,\delta{\bf X},
\end{equation}
where
\begin{equation}
J=
\frac{\partial(x',y',b',u')}
{\partial(x,y,b,u)}
\end{equation}
is the full four-dimensional Jacobian matrix. For a hyperbolic point, $\mathrm{Re}(\lambda_i)<0$ for all eigenvalues
corresponds to a stable attractor, while eigenvalues with both positive and
negative real parts identify a saddle. A positive spectrum corresponds to an
unstable source. The presence of a vanishing real part indicates a
non-hyperbolic configuration for which linear stability alone is generally
insufficient.

For compactness, we define
\begin{equation}
S\equiv
\sqrt{4\gamma_0+\omega_{m0}^{\,2}}.
\label{eq:Sdef}
\end{equation}
The nontrivial branches $P_7$ and $P_8$ are real provided
\begin{equation}
4\gamma_0+\omega_{m0}^{\,2}\geq0,
\end{equation}
with their respective denominators nonzero. For the particle-creation branch
$\gamma_0>0$, one has $S>|\omega_{m0}|$.

For generic parameter values, the named critical configurations and their
Jacobian eigenvalues are summarized in Table~\ref{tab:critical}. The
non-isolated matter critical line at $u=0$ is discussed separately below.

\begin{table}[htbp]
\centering
\scriptsize
\renewcommand{\arraystretch}{1.2}
\begin{tabular}{p{0.75cm} p{3.3cm} p{10.3cm}}
\toprule
Point & Coordinates $(x,y,b,u)$ & Eigenvalues $\{\lambda_i\}$ \\
\midrule

$P_1$
& $(0,0,1,0)$
& $\left\{\frac32,-1,0,-3\omega_{\rm de}\right\}$ \\

$P_2$
& $(0,1,0,0)$
& $\left\{2,1,1,1-3\omega_{\rm de}\right\}$ \\

$P_3$
& $(1,0,0,0)$
& $\left\{\frac32,-1,0,-3\omega_{\rm de}\right\}$ \\

$P_4$
& $(0,0,1,1)$
& $\left\{-\frac32,-1,
12\gamma_0-6\omega_{m0},-3\omega_{\rm de}\right\}$ \\

$P_5$
& $(0,1,0,1)$
& $\left\{-2,1,
1+12\gamma_0-6\omega_{m0},
1-3\omega_{\rm de}\right\}$ \\

$P_6$
& $(1,0,0,1)$
& $\left\{
-1-12\gamma_0+6\omega_{m0},
-\frac32+6\gamma_0-3\omega_{m0},
6(\omega_{m0}-2\gamma_0),
-3(4\gamma_0-2\omega_{m0}+\omega_{\rm de})
\right\}$ \\

$P_7$
& $\left(1,0,0,-\dfrac{1}{\omega_{m0}+S}\right)$
& $\left\{
-4,-3,-3(1+\omega_{\rm de}),
-\dfrac{3S(S+\omega_{m0}+1)}
{(S+\omega_{m0})^2}
\right\}$ \\

$P_8$
& $\left(1,0,0,\dfrac{1}{S-\omega_{m0}}\right)$
& $\left\{
-4,-3,-3(1+\omega_{\rm de}),
-\dfrac{3S(S-\omega_{m0}-1)}
{(S-\omega_{m0})^2}
\right\}$ \\

$P_9$
& $(0,0,0,0)$
& $\left\{
3\omega_{\rm de},
3\omega_{\rm de},
\frac32(1+\omega_{\rm de}),
-1+3\omega_{\rm de}
\right\}$ \\

$P_{10}$
& $(0,0,0,1)$
& $\left\{
3\omega_{\rm de},
-\frac32(1+\omega_{\rm de}),
3(4\gamma_0-2\omega_{m0}+\omega_{\rm de}),
-1+3\omega_{\rm de}
\right\}$ \\

\bottomrule
\end{tabular}
\caption{Critical configurations and eigenvalues of the four-dimensional
autonomous system for generic parameter values. The points $P_1$ and $P_3$
are the endpoints of the continuous matter critical line $L_M$.}
\label{tab:critical}
\end{table}

At $u=0$, the effective DM equation of state satisfies
$\omega_{m,\rm eff}=0$. Baryons and DM therefore have identical background
scaling, giving rise to the continuous matter critical line
\begin{equation}
L_M=(x,0,1-x,0),
\qquad
0\leq x\leq1.
\label{eq:matter_line}
\end{equation}
Its endpoints are $P_1$ and $P_3$. The Jacobian spectrum along this line
contains a zero eigenvalue,
\begin{equation}
\left\{
\frac32,-1,-3\omega_{\rm de},0
\right\},
\end{equation}
reflecting the continuous degeneracy between the baryonic and DM
contributions. Since the spectrum contains both $3/2$ and $-1$, the matter
line is non-hyperbolic and of saddle type.

The radiation point $P_2$ is a source for
$\omega_{\rm de}<1/3$, while it becomes a saddle for
$\omega_{\rm de}>1/3$; the limiting case
$\omega_{\rm de}=1/3$ is non-hyperbolic. The point $P_5$ is always a
saddle because its spectrum contains both $-2$ and $+1$.

The point $P_4$ is linearly stable when
\begin{equation}
\omega_{m0}>2\gamma_0,
\qquad
\omega_{\rm de}>0.
\label{eq:P4_stability}
\end{equation}
These conditions are not relevant to the observationally preferred
negative-$\omega_{\rm de}$ regime, but they characterize the mathematical
stability of the point.

For $P_6$, it is convenient to define
\begin{equation}
d\equiv\omega_{m0}-2\gamma_0.
\end{equation}
Requiring all four eigenvalues to have negative real parts gives
\begin{equation}
-\frac12<d<0,
\qquad
\omega_{\rm de}>2d.
\label{eq:P6_stability}
\end{equation}

For the $P_7$ branch, the general hyperbolic stability conditions are
\begin{equation}
\omega_{\rm de}>-1,
\qquad
S+\omega_{m0}+1>0.
\label{eq:P7_stability}
\end{equation}
For $\gamma_0>0$, however, $S>|\omega_{m0}|$, so that the second condition
is automatically satisfied. In that case, the mathematical $P_7$ branch has
a negative eigenvalue spectrum whenever $\omega_{\rm de}>-1$. Its
cosmological viability must nevertheless be assessed independently from its
location in the compact physical phase space.

For $P_8$, all eigenvalues are negative when
\begin{equation}
S-\omega_{m0}>1,
\qquad
\omega_{\rm de}>-1.
\label{eq:P8_stability}
\end{equation}
Since
\begin{equation}
u_8=\frac{1}{S-\omega_{m0}},
\end{equation}
the strict stability condition $S-\omega_{m0}>1$ simultaneously implies
$0<u_8<1$. The boundary case $S-\omega_{m0}=1$ gives $u_8=1$ and a
vanishing fourth eigenvalue, and is therefore non-hyperbolic.

For $-1<\omega_{\rm de}<0$, the point $P_9$ is necessarily a saddle,
because its spectrum contains the positive eigenvalue
$\frac32(1+\omega_{\rm de})$.

In contrast, $P_{10}$ is linearly stable when
\begin{equation}
-1<\omega_{\rm de}<0,
\qquad
4\gamma_0-2\omega_{m0}+\omega_{\rm de}<0.
\label{eq:P10_stability}
\end{equation}
Since $P_{10}$ is dark-energy dominated, accelerated expansion additionally
requires $\omega_{\rm de}<-1/3$. Consequently, a stable and accelerating
$P_{10}$ endpoint satisfies
\begin{equation}
-1<\omega_{\rm de}<-\frac13,
\qquad
4\gamma_0-2\omega_{m0}+\omega_{\rm de}<0.
\label{eq:P10_stable_acc}
\end{equation}

The eigenvalue analysis establishes the local dynamical character of these
solutions. Their cosmological relevance additionally requires that they lie
inside the compact physical phase space and possess an appropriate cosmic
composition and expansion behavior, which we examine in the following
subsections.

\subsection{Physical properties and cosmological interpretation}
\label{subsec:physical_properties}

The eigenvalue analysis determines the local dynamical character of the
critical configurations, but a cosmologically viable solution must also lie
inside the physical phase space and possess an appropriate matter content
and expansion history. The relevant quantities are the total matter fraction
$\Omega_M=\Omega_{\rm DM}+\Omega_b$, the dark-energy fraction, the effective
DM equation of state, the total equation of state, and the deceleration
parameter. Their values at the critical configurations are summarized in
Table~\ref{tab:physical-points}.

\begin{table}[htbp]
\centering
\renewcommand{\arraystretch}{1.15}
\begin{tabular}{c c c c c c}
\toprule
Point & $\Omega_M$ & $\Omega_{\rm de}$ &
$\omega_{m,\rm eff}$ & $\omega_{\rm tot}$ & $q$ \\
\midrule
$P_1$ & $1$ & $0$ & $0$ & $0$ & $1/2$ \\
$P_2$ & $0$ & $0$ & $0$ & $1/3$ & $1$ \\
$P_3$ & $1$ & $0$ & $0$ & $0$ & $1/2$ \\
$P_4$ & $1$ & $0$ & $2\omega_{m0}-4\gamma_0$ & $0$ & $1/2$ \\
$P_5$ & $0$ & $0$ & $2\omega_{m0}-4\gamma_0$ & $1/3$ & $1$ \\
$P_6$ & $1$ & $0$ &
$2\omega_{m0}-4\gamma_0$ &
$2\omega_{m0}-4\gamma_0$ &
$\frac12[1+3(2\omega_{m0}-4\gamma_0)]$ \\
$P_7$ & $1$ & $0$ & $-1$ & $-1$ & $-1$ \\
$P_8$ & $1$ & $0$ & $-1$ & $-1$ & $-1$ \\
$P_9$ & $0$ & $1$ & $0$ & $\omega_{\rm de}$ &
$\frac12(1+3\omega_{\rm de})$ \\
$P_{10}$ & $0$ & $1$ &
$2\omega_{m0}-4\gamma_0$ &
$\omega_{\rm de}$ &
$\frac12(1+3\omega_{\rm de})$ \\
\bottomrule
\end{tabular}
\caption{Physical quantities evaluated at the critical configurations. For
solutions with $\Omega_{\rm DM}=0$, the formal value of
$\omega_{m,\rm eff}$ does not contribute to $\omega_{\rm tot}$.}
\label{tab:physical-points}
\end{table}

Accelerated expansion occurs when
\begin{equation}
q<0
\qquad\Longleftrightarrow\qquad
\omega_{\rm tot}<-\frac13.
\label{eq:acceleration_condition}
\end{equation}
The points $P_1$--$P_5$ are therefore non-accelerating. The DM-dominated
point $P_6$ can accelerate when
\begin{equation}
2\omega_{m0}-4\gamma_0<-\frac13,
\end{equation}
although accelerated expansion alone does not guarantee that the point is a
stable late-time attractor.

Particular care is required for the nontrivial DM branches $P_7$ and $P_8$,
whose compact coordinates are
\begin{equation}
u_7=-\frac{1}{\omega_{m0}+S},
\qquad
u_8=\frac{1}{S-\omega_{m0}}.
\label{eq:u7_u8}
\end{equation}
For $\gamma_0>0$, one has $S>|\omega_{m0}|$, and therefore
\begin{equation}
u_7<0.
\end{equation}
Hence $P_7$ lies outside the physical compact interval
$0\leq u\leq1$, irrespective of its mathematical stability within the
extended phase space.

For the same particle-creation regime $\gamma_0>0$, the denominator of
$u_8$ is positive, and physical admissibility requires
\begin{equation}
0<u_8\leq1
\qquad\Longleftrightarrow\qquad
S-\omega_{m0}\geq1.
\label{eq:P8_physical}
\end{equation}
The strict inequality $S-\omega_{m0}>1$ corresponds to an interior point
and coincides with the condition for its fourth eigenvalue to be negative.
At the boundary $S-\omega_{m0}=1$, one has $u_8=1$; the fourth eigenvalue
vanishes, and $P_8$ becomes non-hyperbolic and merges with $P_6$. Thus,
although both $P_7$ and $P_8$ formally satisfy
$\omega_{\rm tot}=-1$ and $q=-1$, their cosmological viability cannot be
inferred from their effective equation of state alone.

The points $P_9$ and $P_{10}$ are dark-energy dominated,
\begin{equation}
\Omega_M=0,
\qquad
\Omega_{\rm de}=1,
\qquad
\omega_{\rm tot}=\omega_{\rm de},
\end{equation}
and accelerate whenever
\begin{equation}
\omega_{\rm de}<-\frac13.
\end{equation}
For $-1<\omega_{\rm de}<0$, $P_9$ is always a saddle; it is specifically an
accelerating saddle in the interval
\begin{equation}
-1<\omega_{\rm de}<-\frac13.
\end{equation}
By contrast, $P_{10}$ can simultaneously be accelerating and linearly stable
when
\begin{equation}
-1<\omega_{\rm de}<-\frac13,
\qquad
4\gamma_0-2\omega_{m0}+\omega_{\rm de}<0,
\end{equation}
as summarized in Eq.~\eqref{eq:P10_stable_acc}. Since $P_{10}$ lies at the compact boundary $u=1$, the relation
\begin{equation}
\frac{H}{H_0}=\frac{1-u}{u}
\end{equation}
implies
\begin{equation}
H\rightarrow0
\qquad
\text{as}
\qquad
u\rightarrow1.
\end{equation}
Therefore, for the observationally relevant quintessence-like range
$-1<\omega_{\rm de}<-1/3$, $P_{10}$ represents a compactified,
dark-energy-dominated accelerating endpoint rather than a finite-$H$
de Sitter solution.

The observational analysis presented below determines which of these
dynamically allowed configurations is selected by the inferred parameter
region. 
 
\section{Observational Data, Cosmological Model Framework, and Statistical Methodology}
\label{sec:observations}

\subsection{Observational datasets}
\label{subsec:data}

We constrain the particle-creation cosmology using late-time probes of the background expansion history. This choice is motivated by the phenomenological nature of the model, which is constructed primarily as an effective description of the late-time dark sector. 

\textit{Baryon acoustic oscillations:}
We use the first-year DESI BAO measurements~\cite{DESI:2024mwx,DESI:2024jxi}, building on the established BAO standard-ruler methodology \cite{SDSS:2005xqv,Gaztanaga:2008xz,Oka:2013cba}, covering approximately $0.1\lesssim z\lesssim4.2$. The BAO information is incorporated through the compressed observables
$D_M(z)/r_d$, $D_H(z)/r_d$, and $D_V(z)/r_d$, as appropriate for
the individual redshift bins. Here $D_M(z)$ is the transverse
comoving distance, $D_H(z)=c/H(z)$ is the Hubble distance,
$D_V(z)=[zD_M^2(z)D_H(z)]^{1/3}$ is the volume-averaged
distance, and $r_d$ denotes the sound horizon at the
baryon-drag epoch.  Since the present analysis is intentionally formulated without imposing an early-Universe calibration of $r_d$, we treat the combination $h r_d$ as a free phenomenological parameter. This minimizes the dependence of the late-time constraints on recombination physics and early-Universe baryon-density assumptions. Related late-time analyses using BAO and expansion-history information illustrate the usefulness of treating the low-redshift distance scale with minimal early-Universe assumptions \cite{deCruzPerez:2024shj,Lu:2024hvv,Colgain:2024xqj,Gong:2025hoy,Gao:2025ozb,Gadbail:2024rpp}.

\textit{Type Ia supernovae:}
Three independent Type Ia supernova compilations are considered: Union3~\cite{Rubin:2023jdq}, Pantheon+SH0ES (PPS)~\cite{Scolnic:2021amr,Brout2022}, and the DES-SN5YR Dovekie compilation~\cite{DES:2024jxu,DES:2025sig}. Union3 extends over approximately $0.01\lesssim z\lesssim2.26$, while the PPS sample covers $0.001\lesssim z\lesssim2.26$. For the DES analysis we employ the DES-Dovekie sample constructed within the SALT3.DES5YR framework, containing 1820 supernovae over approximately $0.01\lesssim z\lesssim1.13$. For each compilation, the likelihood is evaluated using the corresponding covariance matrix supplied with the data, with the supernova absolute-magnitude nuisance parameter analytically marginalized.

\textit{Observational Hubble data:}
Direct measurements of the expansion rate are included through the cosmic-chronometer method~\cite{Moresco2016,Moresco2012,Jimenez2003,Simon2005}, which infers $H(z)$ from the differential age evolution of passively evolving galaxies. We use 32 cosmic-chronometer  measurements~\cite{Stern2010,Moresco2015,Ratsimbazafy2017,Borghi2022} spanning approximately $0.07\lesssim z\lesssim2.36$. These measurements provide a direct constraint on the late-time expansion rate without requiring a distance-ladder calibration.

\subsection{Cosmological parameterization and statistical methodology}
\label{subsec:cosmo_parameters}

We constrain the model using the six-dimensional parameter space
\begin{equation}
\boldsymbol{\theta}
=
\left(
\Omega_{m0},h,hr_d,\omega_{\rm de},
\omega_{m,\rm eff}^{0},\gamma_0
\right),
\label{eq:parameter_basis}
\end{equation}
where $\Omega_{m0}$ is the present matter density parameter,
$h$ is the reduced Hubble constant, $hr_d$ is the free BAO
distance-scale parameter, $\omega_{\rm de}$ is the dark-energy
equation-of-state parameter, $\omega_{m,\rm eff}^{0}$ is the
present effective DM equation-of-state parameter, and $\gamma_0$
characterizes particle creation. The intrinsic DM parameter
entering the dynamical system is reconstructed as $
\omega_{m0}
=
\omega_{m,\rm eff}^{0}+\gamma_0$,
and is therefore not an additional free parameter.

Parameter estimation is performed using the affine-invariant
MCMC sampler \texttt{emcee} \cite{ForemanMackey:2013}.
Assuming statistical independence among the observational probes,
the combined likelihood is $
\mathcal{L}_{\rm tot}
=
\mathcal{L}_{\rm SN}
\mathcal{L}_{\rm BAO}
\mathcal{L}_{\rm OHD}$, 
with
$
\chi_{\rm tot}^{2}
=
\chi_{\rm SN}^{2}
+
\chi_{\rm BAO}^{2}
+
\chi_{\rm OHD}^{2}$. The three supernova compilations are analyzed separately,
yielding PPS+BAO+OHD, Union3+BAO+OHD, and
DES-DOV+BAO+OHD. Marginalized posterior constraints are obtained
using standard MCMC post-processing tools \cite{Lewis:2019xzd}.
The resulting chains are also used to reconstruct $\omega_{m0}$
and examine the stability conditions of the cosmological
critical points. 

\begin{table}[htbp]
\scriptsize
\centering
\renewcommand{\arraystretch}{1.25}
\begin{tabular}{lccc}
\toprule
Parameter &
PPS+BAO+OHD &
Union3+BAO+OHD &
DES-DOV+BAO+OHD \\
\midrule

$\Omega_{m0}$ &
$0.3031^{+0.0196}_{-0.0197}$ &
$0.3030^{+0.0197}_{-0.0199}$ &
$0.3026^{+0.0197}_{-0.0196}$ \\

$h$ &
$0.6692^{+0.0079}_{-0.0078}$ &
$0.6576^{+0.0100}_{-0.0099}$ &
$0.6684^{+0.0073}_{-0.0073}$ \\

$h r_d$ &
$100.105^{+1.053}_{-1.026}$ &
$98.553^{+1.333}_{-1.304}$ &
$99.996^{+0.958}_{-0.941}$ \\

$\omega_{\rm de}$ &
$-0.8787^{+0.0432}_{-0.0447}$ &
$-0.8110^{+0.0562}_{-0.0575}$ &
$-0.8716^{+0.0401}_{-0.0427}$ \\

$\omega_{m,\rm eff}^{0}$ &
$-0.0535^{+0.0373}_{-0.0399}$ &
$-0.0562^{+0.0395}_{-0.0425}$ &
$-0.0544^{+0.0382}_{-0.0397}$ \\

$\gamma_0$ &
$0.1366^{+0.1012}_{-0.0931}$ &
$0.1525^{+0.0966}_{-0.1020}$ &
$0.1429^{+0.0991}_{-0.0964}$ \\

$\omega_{m0}$ &
$0.0876^{+0.0751}_{-0.0749}$ &
$0.0998^{+0.0687}_{-0.0790}$ &
$0.0923^{+0.0735}_{-0.0759}$ \\

\bottomrule
\end{tabular}
\caption{Marginalized posterior constraints on the cosmological parameters
for the three observational combinations. The quoted values correspond to
the posterior median with the $68\%$ credible interval. The intrinsic
present-day dark-matter equation-of-state parameter is derived
sample-by-sample from
$\omega_{m0}=\omega_{m,\rm eff}^{0}+\gamma_0$, thereby retaining the
posterior correlation between $\omega_{m,\rm eff}^{0}$ and $\gamma_0$, shown in Fig.~\ref{fig:corner}.}
\label{tab:posterior_constraints}
\end{table}

\subsection{Posterior constraints and parameter correlations}
\label{subsec:posterior}

\begin{figure}[H]
    \centering
    \includegraphics[width=0.94\textwidth]{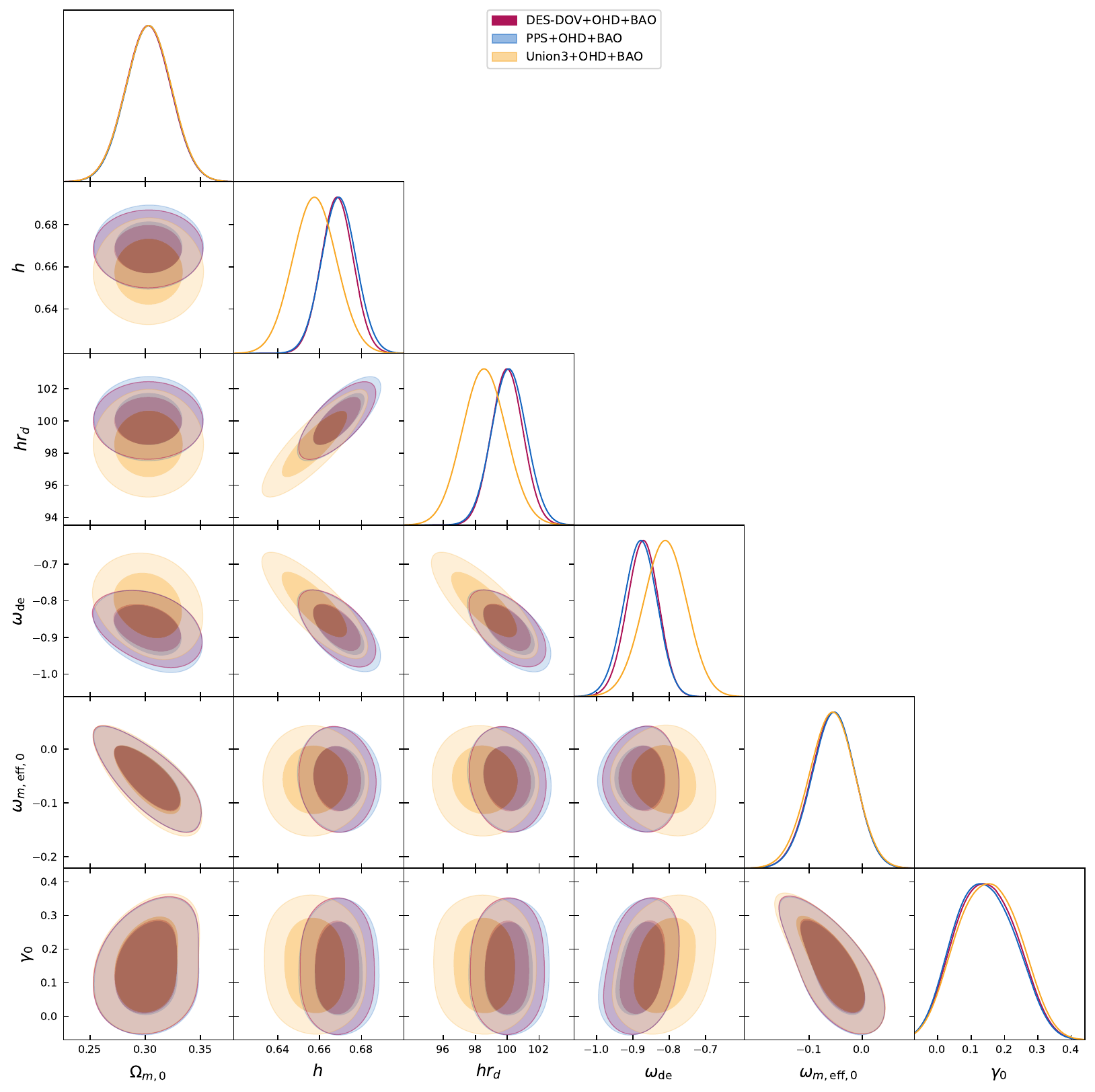}
    \caption{One- and two-dimensional marginalized posterior
    distributions of
    $(\Omega_{m0},h,hr_d,\omega_{\rm de},
    \omega_{m,\rm eff}^{0},\gamma_0)$
    for DES-DOV+BAO+OHD, PPS+BAO+OHD, and
    Union3+BAO+OHD.}
    \label{fig:corner}
\end{figure}

The marginalized posterior distributions for the three observational
combinations are presented in Fig.~\ref{fig:corner}. The inferred
parameters occupy broadly overlapping regions of the common
six-dimensional parameter space, indicating consistency among the
supernova compilations. The most noticeable differences occur in
$h$ and $\omega_{\rm de}$, with Union3 favoring a slightly lower
Hubble parameter and a less negative dark-energy equation of state
than PPS and DES-DOV. All three combinations favor quintessence-like
central values of $\omega_{\rm de}$, mildly negative values of
$\omega_{m,\rm eff}^{0}$, and positive particle-creation amplitudes.
Nevertheless, the relatively broad posterior of $\gamma_0$
indicates that its determination from background observations
remains limited.

\begin{figure}[H]
    \centering
    \includegraphics[width=0.96\textwidth]{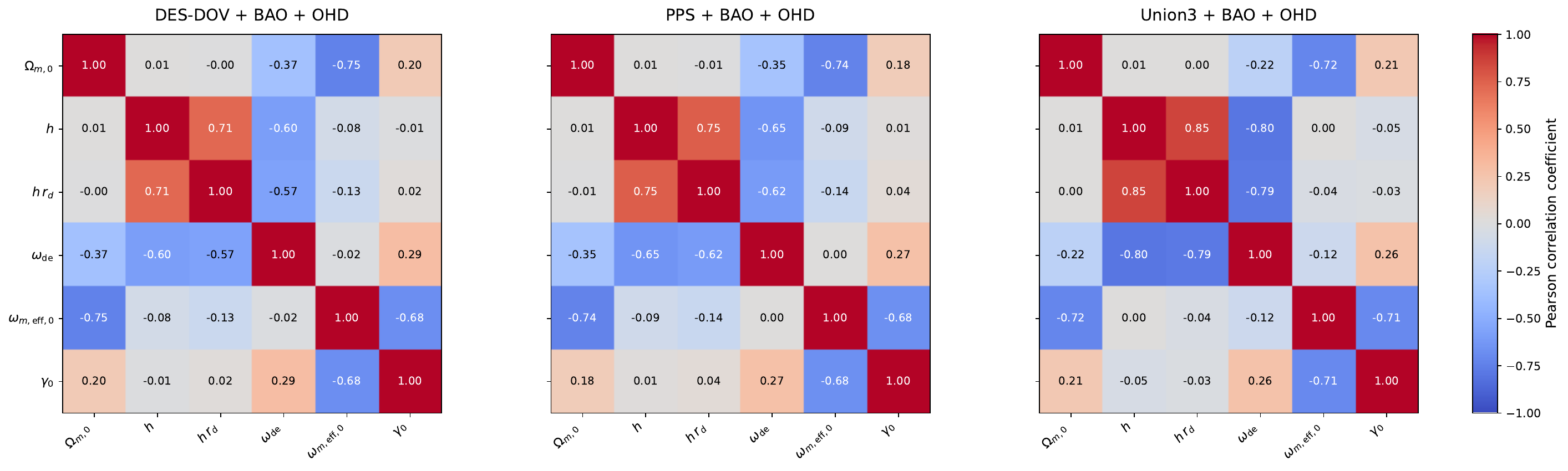}
    \caption{Pearson correlation matrices for the six sampled
    cosmological parameters. The dominant features include
    the positive $h$--$hr_d$ correlation and the negative
    $\omega_{m,\rm eff}^{0}$--$\gamma_0$ correlation.}
    \label{fig:corr-heatmap}
\end{figure}

The Pearson correlation matrices in Fig.~\ref{fig:corr-heatmap}
reveal several parameter degeneracies. In particular, $h$ and
$hr_d$ exhibit a positive correlation, with coefficients of
approximately $0.71$--$0.85$, reflecting the interplay between
the expansion scale and the BAO standard ruler. Both parameters
are anticorrelated with $\omega_{\rm de}$, illustrating the
compensation between changes in the dark-energy expansion history
and cosmological distances.

A pronounced negative correlation is also observed between
$\omega_{m,\rm eff}^{0}$ and $\gamma_0$, with Pearson coefficients
close to $-0.7$. This reflects a degeneracy between the intrinsic
DM equation of state and the particle-creation contribution,
which jointly determine the effective DM evolution through
\begin{equation}
\omega_{m,\rm eff}^{0}
=
\omega_{m0}-\gamma_0.
\end{equation}
The total matter fraction $\Omega_{m0}$ is additionally
anticorrelated with $\omega_{m,\rm eff}^{0}$, indicating that
variations in the effective DM pressure can be partially
compensated by changes in the present matter abundance.

\begin{figure}[!ht]
    \centering

    \begin{subfigure}[t]{0.43\textwidth}
        \centering
        \includegraphics[width=\linewidth]
        {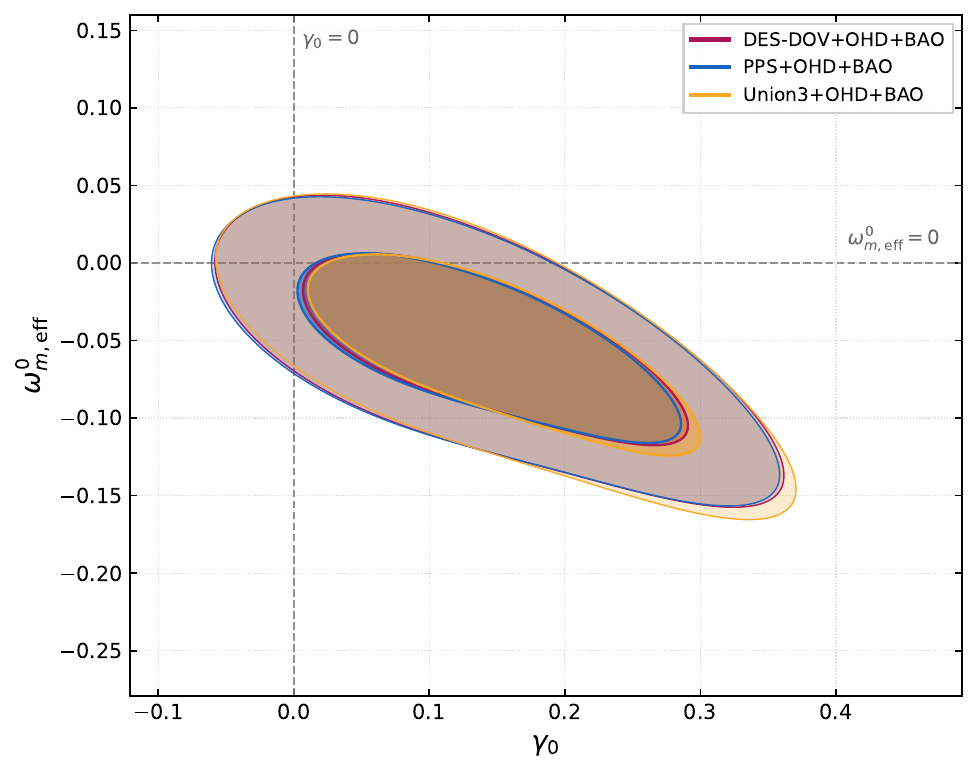}
        \caption{$\gamma_0$ vs.\
        $\omega_{m,\rm eff}^{0}$.}
    \end{subfigure}
    \hspace{0.03\textwidth}
    \begin{subfigure}[t]{0.43\textwidth}
        \centering
        \includegraphics[width=\linewidth]
        {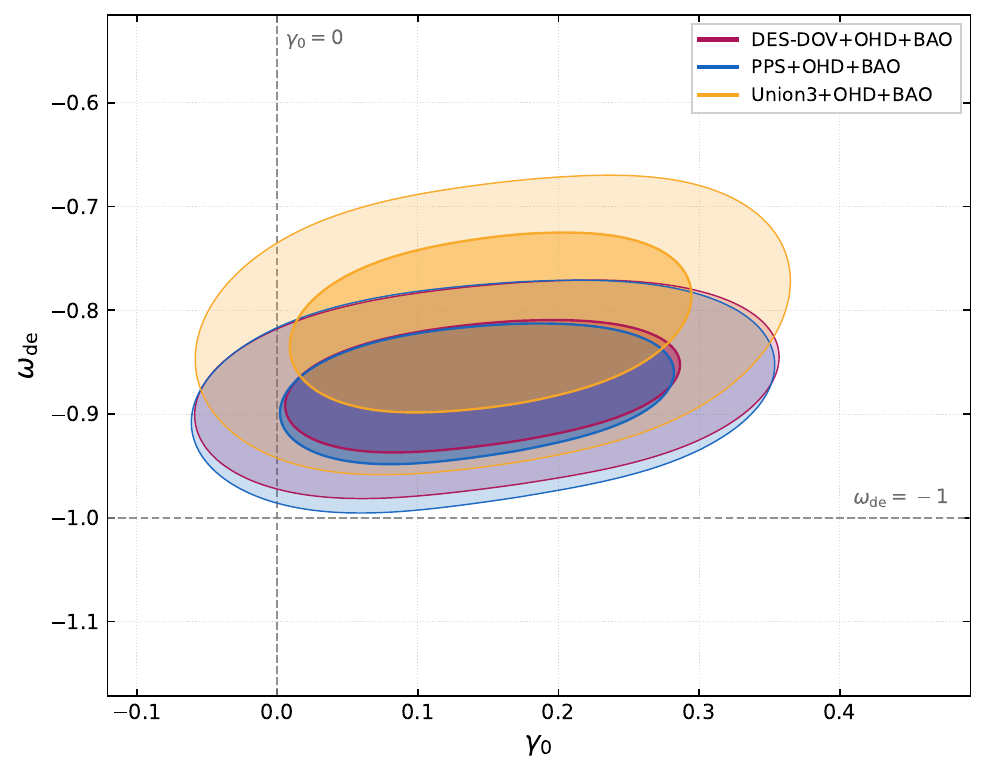}
        \caption{$\gamma_0$ vs.\
        $\omega_{\rm de}$.}
    \end{subfigure}

    \vspace{0.1cm}

    \begin{subfigure}[t]{0.43\textwidth}
        \centering
        \includegraphics[width=\linewidth]
        {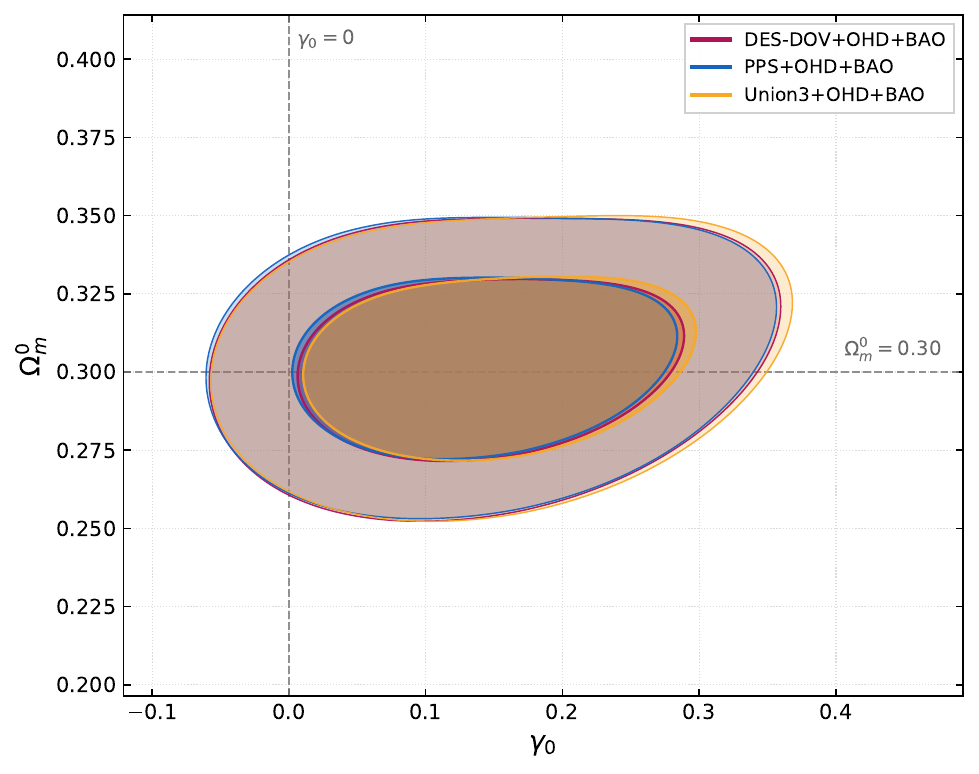}
        \caption{$\gamma_0$ vs.\
        $\Omega_{m0}$.}
    \end{subfigure}
    \hspace{0.03\textwidth}
    \begin{subfigure}[t]{0.43\textwidth}
        \centering
        \includegraphics[width=\linewidth]
        {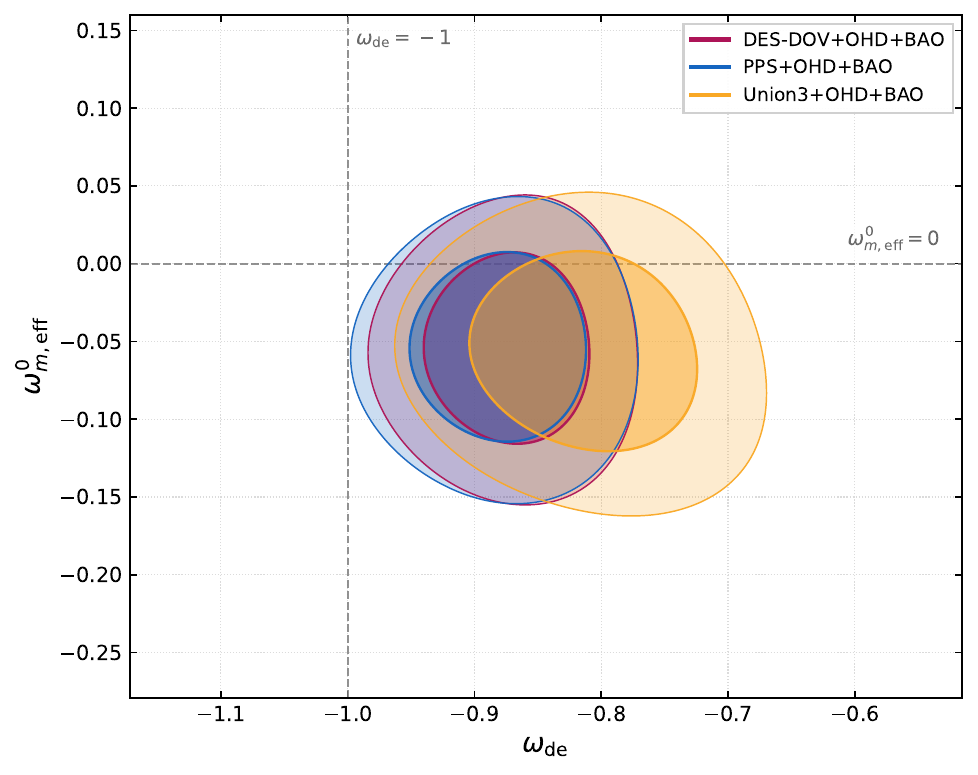}
        \caption{$\omega_{\rm de}$ vs.\
        $\omega_{m,\rm eff}^{0}$.}
    \end{subfigure}

    \caption{Selected two-dimensional marginalized posterior
    constraints for the three observational combinations.
    The inner and outer contours represent the $68\%$
    and $95\%$ credible regions, respectively.
    Dashed reference lines indicate
    $\gamma_0=0$, $\omega_{m,\rm eff}^{0}=0$,
    $\omega_{\rm de}=-1$, and $\Omega_{m0}=0.30$.}
    \label{fig:selected-2d}
\end{figure}

Selected two-dimensional constraints are displayed in
Fig.~\ref{fig:selected-2d}. The
$\gamma_0$--$\omega_{m,\rm eff}^{0}$ contours show a strong
anticorrelation, while the $\gamma_0$--$\omega_{\rm de}$
plane exhibits a substantially weaker, mildly positive
correlation. The $\gamma_0$--$\Omega_{m0}$ contours indicate
that the matter abundance is more tightly constrained than
the particle-creation amplitude, with the preferred matter
fraction remaining close to $\Omega_{m0}\simeq0.30$.
Finally, the $\omega_{\rm de}$--$\omega_{m,\rm eff}^{0}$
constraints show that the principal differences among the
supernova combinations arise in the dark-energy sector,
whereas the inferred effective DM behavior remains broadly
consistent.

\section{Observationally Calibrated Dynamical Behaviour}
\label{sec:obs_dynamics}

The posterior constraints can be implemented directly into the phase-space language developed in Sec.~\ref{sec:dynamical_system}. We first evaluate the critical-point structure at the maximum-likelihood parameter sets and then extend the analysis to the full posterior volume. This separates two logically distinct questions: whether a given best-fit cosmology approaches a particular asymptotic state, and how robust that phase-space classification is under the observational uncertainties.

\subsection{Physicality and stability of the late-time critical points}
\label{subsec:numerical_critical}

\begin{figure*}[htbp]
    \centering
    \includegraphics[width=0.98\textwidth]{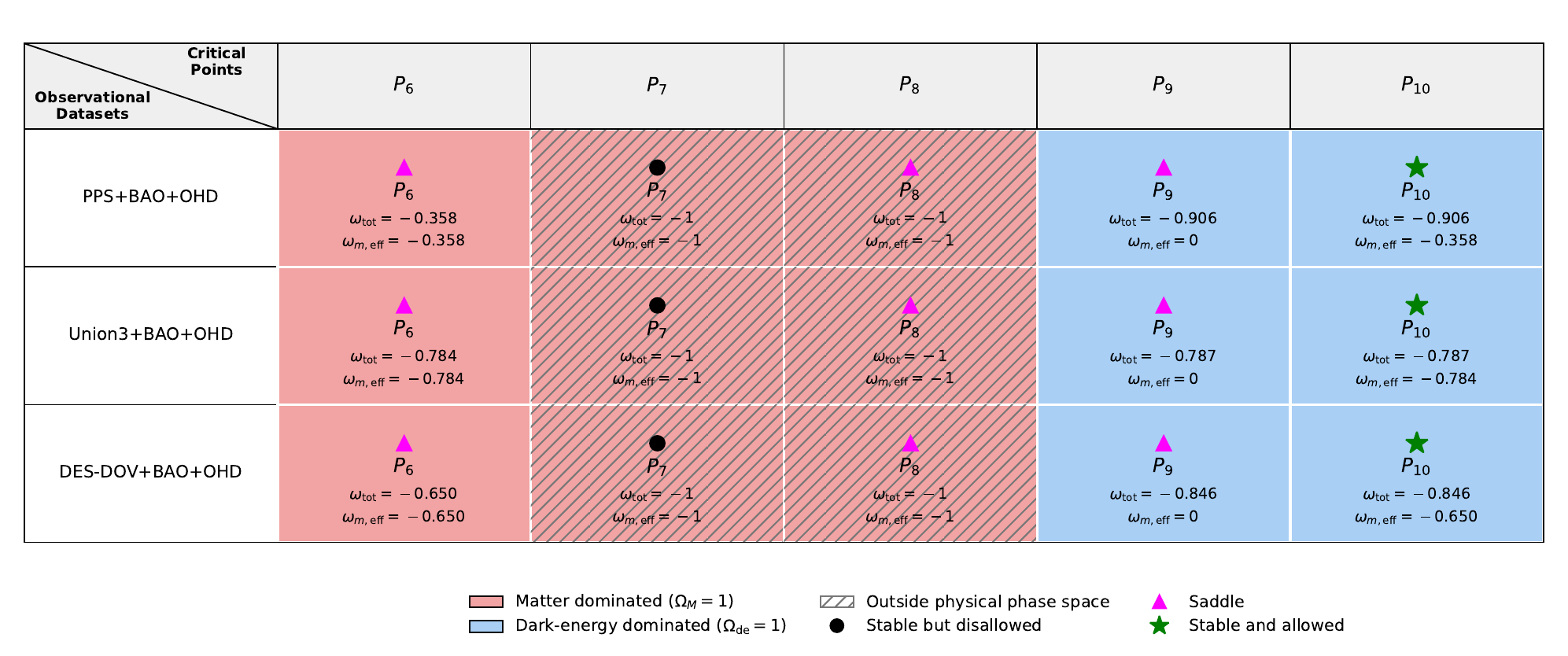}
    \caption{
    Physicality and linear-stability properties of the late-time critical
    points $P_6$--$P_{10}$ evaluated at the three observational best-fit
    parameter sets. Red cells denote matter-dominated configurations, while
    blue cells correspond to dark-energy-dominated states. Hatched cells mark
    critical points lying outside the physical compact domain $0\leq u\leq1$.
    Black circles indicate stable points, red triangles denote saddle points,
    and the green star highlights the physically allowed stable point $P_{10}$.
    The corresponding values of $\omega_{\rm tot}$ and
    $\omega_{m,\rm eff}$ are displayed in each cell. For the
    dark-energy-dominated points, where $\Omega_{\rm DM}=0$,
    $\omega_{m,\rm eff}$ is dynamically irrelevant to the background
    evolution.
    }
    \label{fig:critical_point_summary}
\end{figure*}

The physicality and stability properties of the DM- and
dark-energy-dominated critical points $P_6$--$P_{10}$ are
summarized in Fig.~\ref{fig:critical_point_summary} for
the three observational best-fit parameter sets. The color coding identifies the dominant background component,
while the symbols indicate the corresponding linear stability character. This
representation emphasizes that accelerated expansion alone does not determine
the physical late-time endpoint. A viable asymptotic state must also lie within
the compact physical domain, $0\leq u\leq1$, and satisfy the appropriate
Jacobian stability conditions.

The matter-dominated point $P_6$ has $\Omega_M=1$ and
\begin{equation}
\omega_{\rm tot}
=
\omega_{m,\rm eff}
=
2\omega_{m0}-4\gamma_0 .
\end{equation}
For all three best-fit parameter sets, this quantity is below $-1/3$, so
$P_6$ corresponds formally to an accelerating matter-dominated configuration.
However, at least one eigenvalue remains positive in each case, and the point is
therefore a saddle rather than a stable future attractor. Thus, although $P_6$
satisfies the kinematical acceleration condition, it does not represent the
dynamically preferred asymptotic state.

The nontrivial DM points $P_7$ and $P_8$ require an additional physicality
check through their compact coordinates,
\begin{equation}
u_7
=
-\frac{1}{\omega_{m0}+S},
\qquad
u_8
=
\frac{1}{S-\omega_{m0}},
\qquad
S
=
\sqrt{4\gamma_0+\omega_{m0}^{2}} .
\end{equation}
For $\gamma_0>0$, one always has $u_7<0$, placing $P_7$ outside the physical
expanding phase space, irrespective of its formal linear stability. For the
three exact best-fit parameter sets, $P_8$ instead satisfies $u_8>1$ and hence
lies beyond the upper compact boundary. In addition, its fourth eigenvalue is
positive for the calibrated best-fit values. Consequently, neither $P_7$ nor
$P_8$ constitutes a physically admissible late-time endpoint for the
observationally preferred trajectories, despite the formal values
\begin{equation}
\omega_{\rm tot}
=
\omega_{m,\rm eff}
=
-1 .
\end{equation}

The remaining points $P_9$ and $P_{10}$ are dark-energy dominated,
\begin{equation}
\Omega_M=0,
\qquad
\Omega_{\rm de}=1,
\qquad
\omega_{\rm tot}=\omega_{\rm de}.
\end{equation}
Since the DM density vanishes at these points, the value of
$\omega_{m,\rm eff}$ displayed in Fig.~\ref{fig:critical_point_summary} is
formally defined but does not contribute to the background dynamics. In the
observationally relevant range $-1<\omega_{\rm de}<0$, $P_9$ can correspond to
accelerated expansion when $\omega_{\rm de}<-1/3$, but it remains a saddle
because of the positive eigenvalue
$\tfrac{3}{2}(1+\omega_{\rm de})$. By contrast, $P_{10}$ satisfies the full
linear-stability conditions for all three exact best-fit parameter sets and
provides the physically allowed stable late-time endpoint shown in
Fig.~\ref{fig:critical_point_summary}. Since $\omega_{\rm de}>-1$ for the
best-fit solutions, this asymptotic state is quintessence-like rather than an
exact de Sitter configuration.

 At the physically allowed late-time point
\[
P_{10}=(0,0,0,1),
\]
the matter contribution vanishes asymptotically, so that
\[
\Omega_M=0,
\qquad
\Omega_{\rm de}=1,
\qquad
\omega_{\rm tot}=\omega_{\rm de}.
\]
The corresponding deceleration parameter is
\[
q_{P_{10}}
=
\frac{1}{2}\left(1+3\omega_{\rm de}\right),
\]
which, for the three exact best-fit parameter sets, gives
\[
q_{P_{10}}^{\rm PPS}=-0.859183,
\qquad
q_{P_{10}}^{\rm Union3}=-0.680058,
\qquad
q_{P_{10}}^{\rm DES-DOV}=-0.768946.
\]
Hence, all three calibrated realizations approach an accelerating
dark-energy-dominated state. The linear-stability requirements,
\[
-1<\omega_{\rm de}<0,
\qquad
4\gamma_0-2\omega_{m0}+\omega_{\rm de}<0,
\]
are also satisfied by the three best-fit parameter sets, confirming that
$P_{10}$ is the dynamically stable and physically allowed late-time endpoint.

The compact coordinate $u=1$ corresponds to the asymptotic
limit $H\rightarrow0$, since $H/H_0=(1-u)/u$.
Therefore, the stable point $P_{10}$ does not represent
a finite-$H$ de Sitter solution. For the observationally
preferred quintessence-like regime
$-1<\omega_{\rm de}<-1/3$, the asymptotic expansion follows
\begin{equation}
a(t)\propto
t^{\frac{2}{3(1+\omega_{\rm de})}},
\qquad
H(t)\propto t^{-1}\rightarrow0.
\end{equation}
Thus, the three observationally calibrated models admit
a stable, accelerating, dark-energy-dominated endpoint
characterized by power-law expansion rather than
asymptotic de Sitter evolution.

\subsection{Reconstruction of cosmological parameters}
\label{subsec:background_reconstruction}

For the redshift-space reconstruction, we define
\begin{equation}
E(z)=\frac{H(z)}{H_0},
\qquad
u(z)=\frac{1}{1+E(z)}.
\end{equation}
The intrinsic and effective dark-matter equations of state, together with the
dimensionless particle-creation strength, are then
\begin{align}
\omega_m(z)
&=
\frac{2\omega_{m0}}{1+E(z)},
\\
\frac{\Gamma(z)}{3H(z)}
&=
\frac{4\gamma_0}{[1+E(z)]^2},
\\
\omega_{m,\rm eff}(z)
&=
\frac{2\omega_{m0}}{1+E(z)}
-
\frac{4\gamma_0}{[1+E(z)]^2}.
\label{eq:redshift_dm_quantities}
\end{align}
The DM density evolves according to
\begin{equation}
\frac{d\rho_m}{dz}
=
\frac{3\rho_m}{1+z}
\left[
1+\omega_m(z)-\frac{\Gamma(z)}{3H(z)}
\right].
\label{eq:rho_dm_z}
\end{equation}

\begin{figure}[htbp]
    \centering
    \begin{subfigure}[t]{0.485\textwidth}
        \centering
        \includegraphics[width=\linewidth]{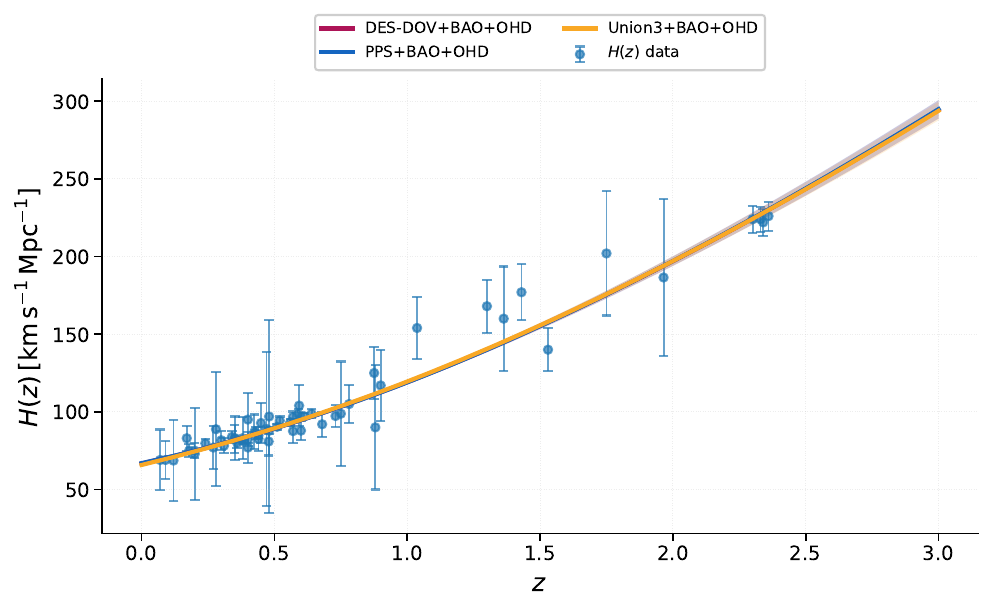}
        \caption{Reconstructed $H(z)$ together with the OHD measurements.}
    \end{subfigure}
    \hfill
    \begin{subfigure}[t]{0.485\textwidth}
        \centering
        \includegraphics[width=\linewidth]{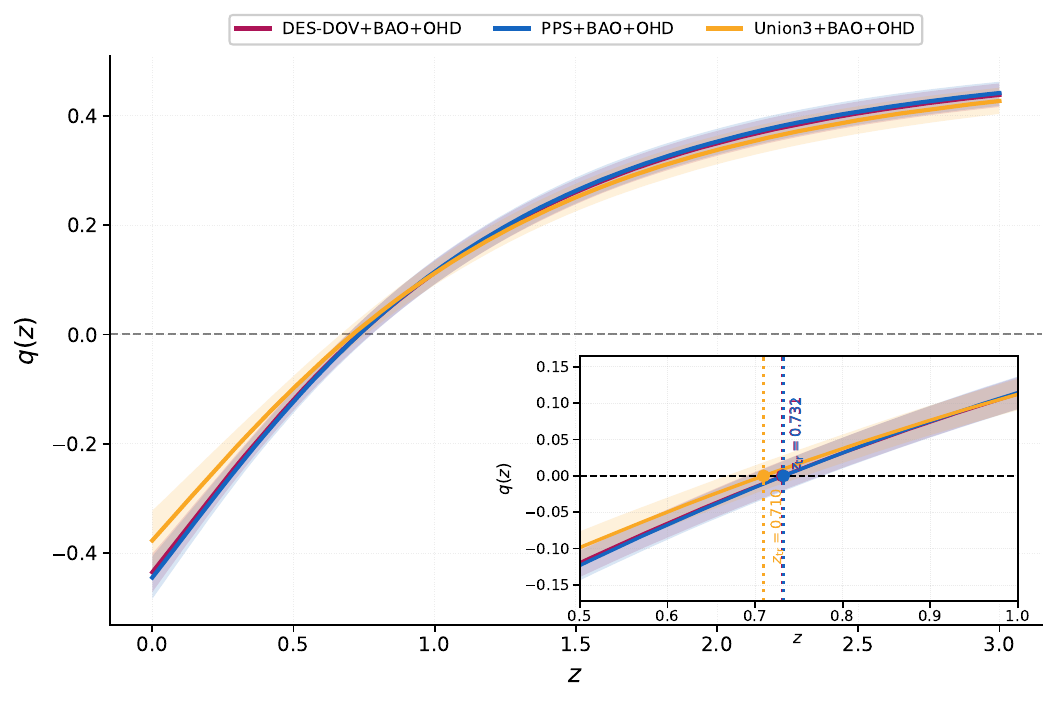}
        \caption{Reconstructed deceleration parameter $q(z)$.}
    \end{subfigure}
    \caption{
    Late-time background reconstruction for the three observational
    combinations. Solid curves denote the posterior median histories and the
    shaded regions the corresponding $68\%$ credible intervals. The zero
    crossing of $q(z)$ marks the transition from decelerated to accelerated
    expansion.
    }
    \label{fig:Hz-qz}
\end{figure}

Figure~\ref{fig:Hz-qz} shows the reconstructed expansion history and the
deceleration parameter. The three $H(z)$ reconstructions are very similar over
the redshift range directly covered by the OHD measurements, indicating that
the different supernova compilations lead to only modest shifts in the
dark-sector parameters while preserving nearly the same background expansion
history. The reconstructed median curves remain consistent with the OHD data
within the quoted uncertainties, while the credible bands broaden in regions
where the chronometer sampling becomes sparse.

The right panel of Fig.~\ref{fig:Hz-qz} provides a direct kinematical measure
of the onset of acceleration. In all three observational combinations,
$q(z)$ evolves from positive values at higher redshift to negative values at
late times. The transition condition,
\begin{equation}
q(z_{\rm tr})=0,
\end{equation}
is reached in the narrow interval
$z_{\rm tr}\simeq0.71$--$0.73$. The close agreement among the three
reconstructions indicates that the transition redshift is robust against the
choice of supernova compilation and is mainly determined by the combined
late-time distance and expansion-rate information.

\begin{figure}[htbp]
    \centering
    \begin{subfigure}[t]{0.485\textwidth}
        \centering
        \includegraphics[width=\linewidth]{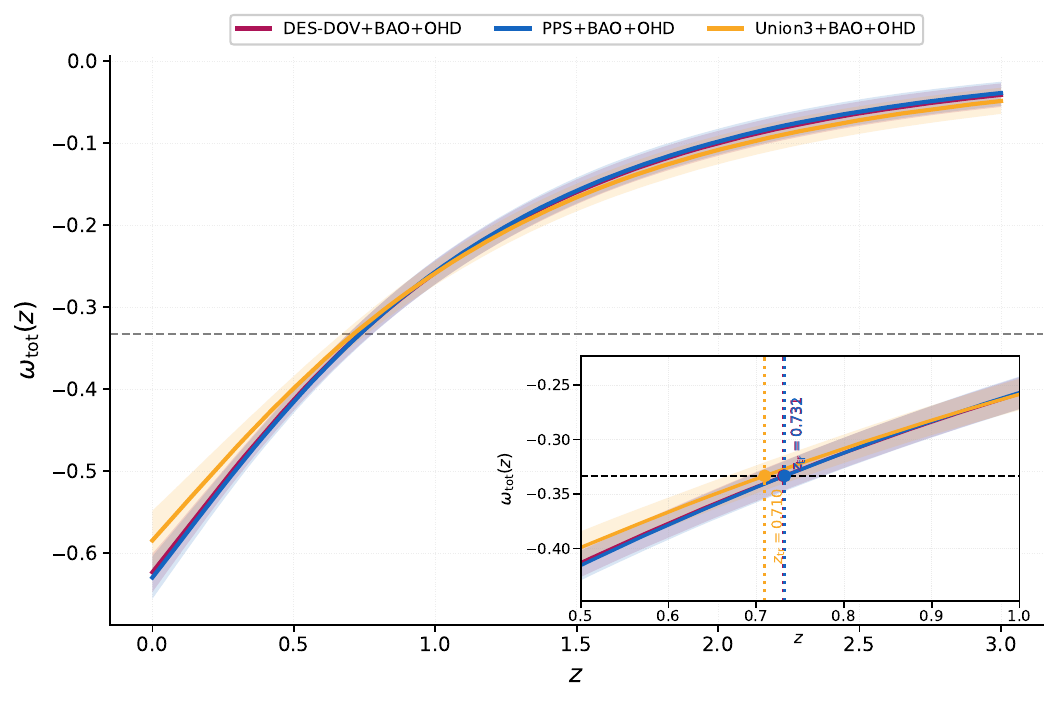}
        \caption{Reconstructed total EoS parameter.}
    \end{subfigure}
    \hfill
    \begin{subfigure}[t]{0.485\textwidth}
        \centering
        \includegraphics[width=\linewidth]{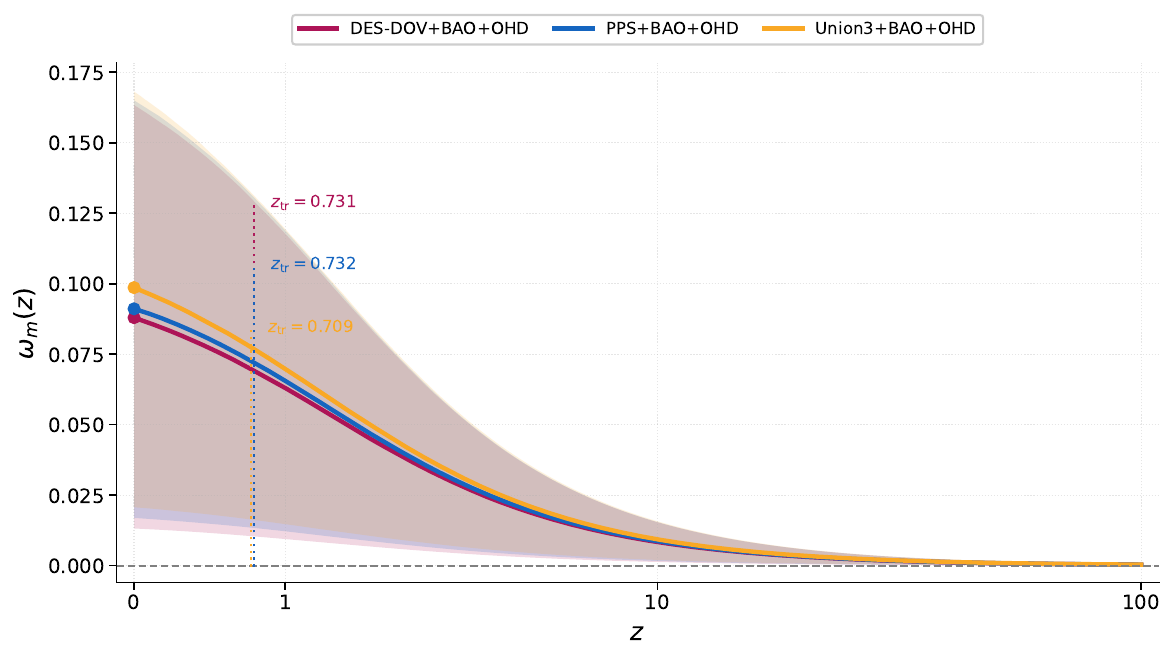}
        \caption{Reconstructed intrinsic DM EoS parameter.}
    \end{subfigure}
    \caption{
    Reconstructed redshift evolution of the equation-of-state parameters.
    The left panel shows the effective equation of state of the total cosmic
    fluid, while the right panel shows the intrinsic dark-matter equation of
    state $\omega_m(z)$. Solid curves denote posterior medians and shaded
    regions the corresponding $68\%$ credible intervals. The vertical
    $z_{\rm tr}$ markers, where shown, denote the acceleration-transition
    redshift obtained from $q(z_{\rm tr})=0$ and are displayed only for
    reference; they do not indicate a transition in $\omega_m(z)$.
    }
    \label{fig:eos-z}
\end{figure}

The reconstructed equations of state are shown in
Fig.~\ref{fig:eos-z}. The total equation of state becomes progressively more
negative toward the present epoch and crosses
$\omega_{\rm tot}=-1/3$ at the same redshift at which $q(z)$ changes sign, as
follows directly from
\begin{equation}
q=\frac{1}{2}\left(1+3\omega_{\rm tot}\right).
\end{equation}
Toward higher redshift, $\omega_{\rm tot}$ approaches the matter-dominated
regime because the dark-energy fraction decreases and the dark-matter
corrections are increasingly suppressed.

The intrinsic DM equation of state, shown in the right panel of
Fig.~\ref{fig:eos-z}, decreases smoothly toward zero at high redshift.
The vertical markers at $z_{\rm tr}$, where shown, indicate the independently
reconstructed transition from decelerated to accelerated expansion,
$q(z_{\rm tr})=0$, and are included only as a kinematical reference; they do
not represent a transition in $\omega_m(z)$ itself. Since
\begin{equation}
\omega_m(z)=2\omega_{m0}u(z),
\end{equation}
and $u\rightarrow0$ in the high-$H$ regime, the intrinsic DM pressure becomes
negligible in the early Universe. Around the present epoch,
$u_0=1/2$ and therefore $\omega_m(0)=\omega_{m0}$, so the low-redshift
behavior directly reflects the reconstructed present-day intrinsic DM
equation of state.

\begin{figure}[H]
    \centering
    \begin{subfigure}[t]{0.485\textwidth}
        \centering
        \includegraphics[width=\linewidth]{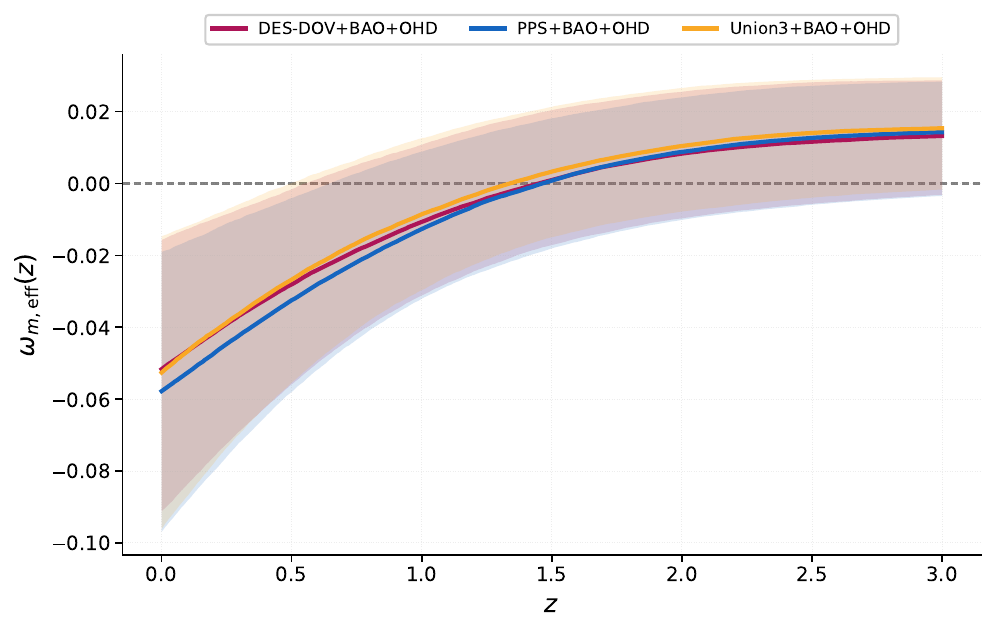}
        \caption{Reconstructed effective dark-matter EoS parameter.}
    \end{subfigure}
    \hfill
    \begin{subfigure}[t]{0.485\textwidth}
        \centering
        \includegraphics[width=\linewidth]{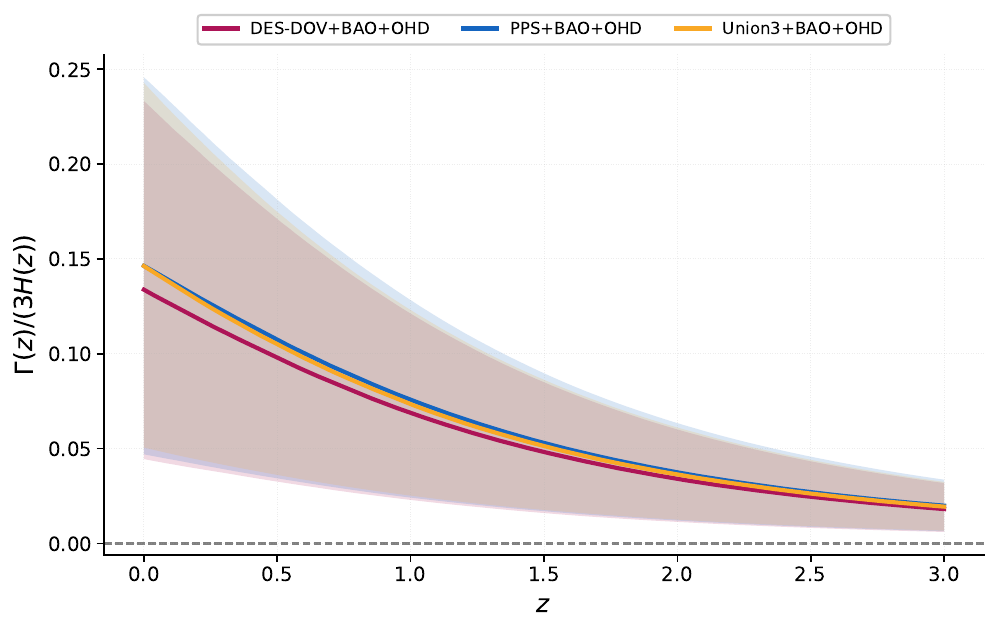}
        \caption{Reconstructed dimensionless creation strength parameter.}
    \end{subfigure}
    \caption{
    Reconstructed evolution of the effective dark-matter equation of state
    and the normalized particle-creation strength. Both quantities are
    suppressed toward higher redshift through their dependence on the compact
    variable $u=H_0/(H_0+H)$, while their larger late-time values reflect the
    increasing relevance of the particle-creation sector as the expansion rate
    decreases.
    }
    \label{fig:creation-z}
\end{figure}
Figure~\ref{fig:creation-z} isolates the effective dark-matter sector and the
particle-creation contribution. Around the present epoch,
$\omega_{m,\rm eff}$ is mildly negative, indicating that the negative creation
pressure slightly exceeds the positive intrinsic DM contribution. As the
redshift increases, both terms are suppressed through their dependence on
$u=1/(1+E)$. The intrinsic contribution is linear in $u$, whereas the
creation contribution is quadratic,
\begin{equation}
\omega_{m,\rm eff}
=
2\omega_{m0}u
-
4\gamma_0u^2.
\end{equation}
Consequently, the creation term decays more rapidly in the high-$H$ regime,
and $\omega_{m,\rm eff}$ approaches the pressureless limit. The small sign
change seen in the reconstructed curves is therefore a direct consequence of
the different $u$-dependences of the intrinsic and creation terms and should
not be interpreted as an independent dynamical transition.

The dimensionless creation strength is
\begin{equation}
\frac{\Gamma}{3H}=4\gamma_0u^2,
\end{equation}
which vanishes quadratically as $u\rightarrow0$. Particle creation is
therefore strongly suppressed at high redshift, and the DM sector approaches
the standard non-creating pressureless limit in the early Universe. The
corresponding dimensional rate is
\begin{equation}
\Gamma
=
12\gamma_0H_0u(1-u),
\end{equation}
which is also suppressed in the high-$H$ past and tends to zero at the compact
future boundary $u\rightarrow1$, where $H\rightarrow0$. The two quantities
should therefore be distinguished carefully: although
$\Gamma\rightarrow0$ as $u\rightarrow1$, the normalized ratio approaches the
finite limit
\begin{equation}
\frac{\Gamma}{3H}
\longrightarrow
4\gamma_0.
\end{equation}

\subsection{Dark-energy significance and posterior stability}
\label{subsec:posterior_dynamics}

While the preceding critical-point analysis relies on the exact
best-fit parameters, the robustness of the late-time attractor
can be assessed across the full posterior distribution.
For each retained MCMC sample, the intrinsic DM parameter is
reconstructed as
$\omega_{m0}=\omega_{m,\rm eff}^{0}+\gamma_0$.
For the nontrivial DM-dominated point $P_8$, we define
$A=\sqrt{4\gamma_0+\omega_{m0}^{2}}-\omega_{m0}$.
This point represents a physically admissible interior solution
when $A>1$, with linear stability additionally requiring
$\omega_{\rm de}>-1$.
For the dark-energy-dominated point $P_{10}$, the stability
condition becomes
$2\gamma_0-2\omega_{m,\rm eff}^{0}+\omega_{\rm de}<0$,
together with $-1<\omega_{\rm de}<-1/3$ for a stable,
accelerating quintessence-like endpoint.
Introducing $X=\omega_{\rm de}$ and
$Y=2(\gamma_0-\omega_{m,\rm eff}^{0})$,
the corresponding stability boundary is $Y=-X$.

\begin{figure}[H]
    \centering
    \begin{subfigure}[t]{0.485\textwidth}
        \centering
        \includegraphics[width=\linewidth]
        {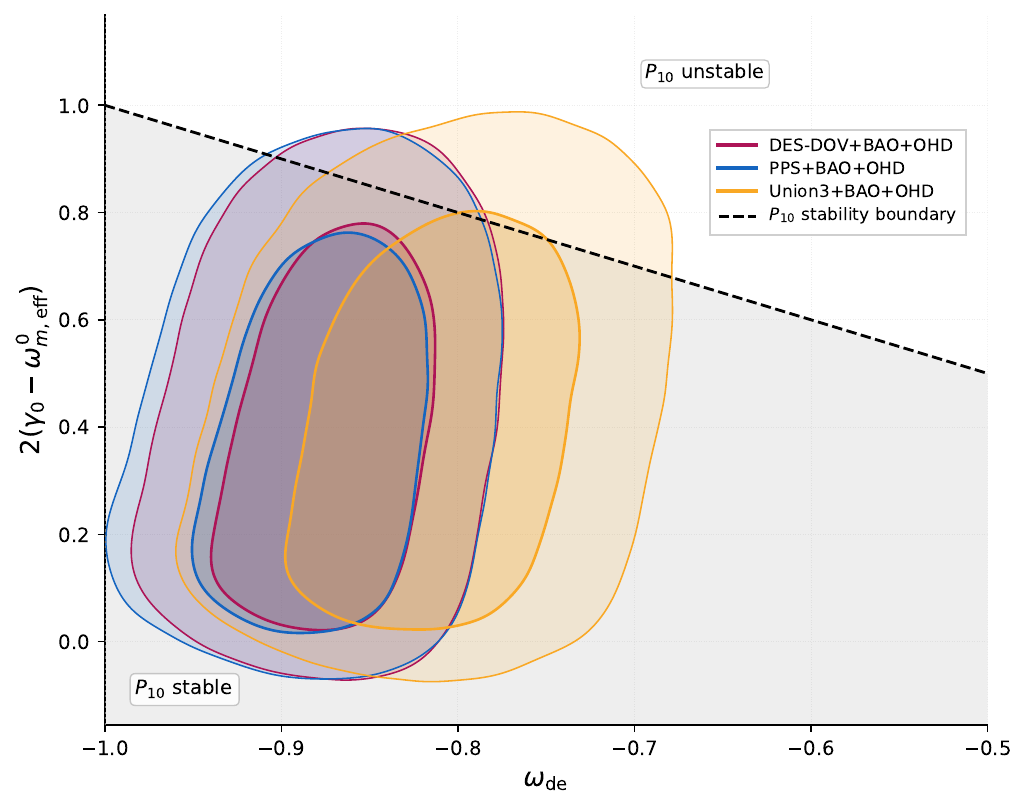}
        \caption{Posterior constraints in the
        $P_{10}$ stability plane.}
    \end{subfigure}
    \hfill
    \begin{subfigure}[t]{0.485\textwidth}
        \centering
        \includegraphics[width=\linewidth]
        {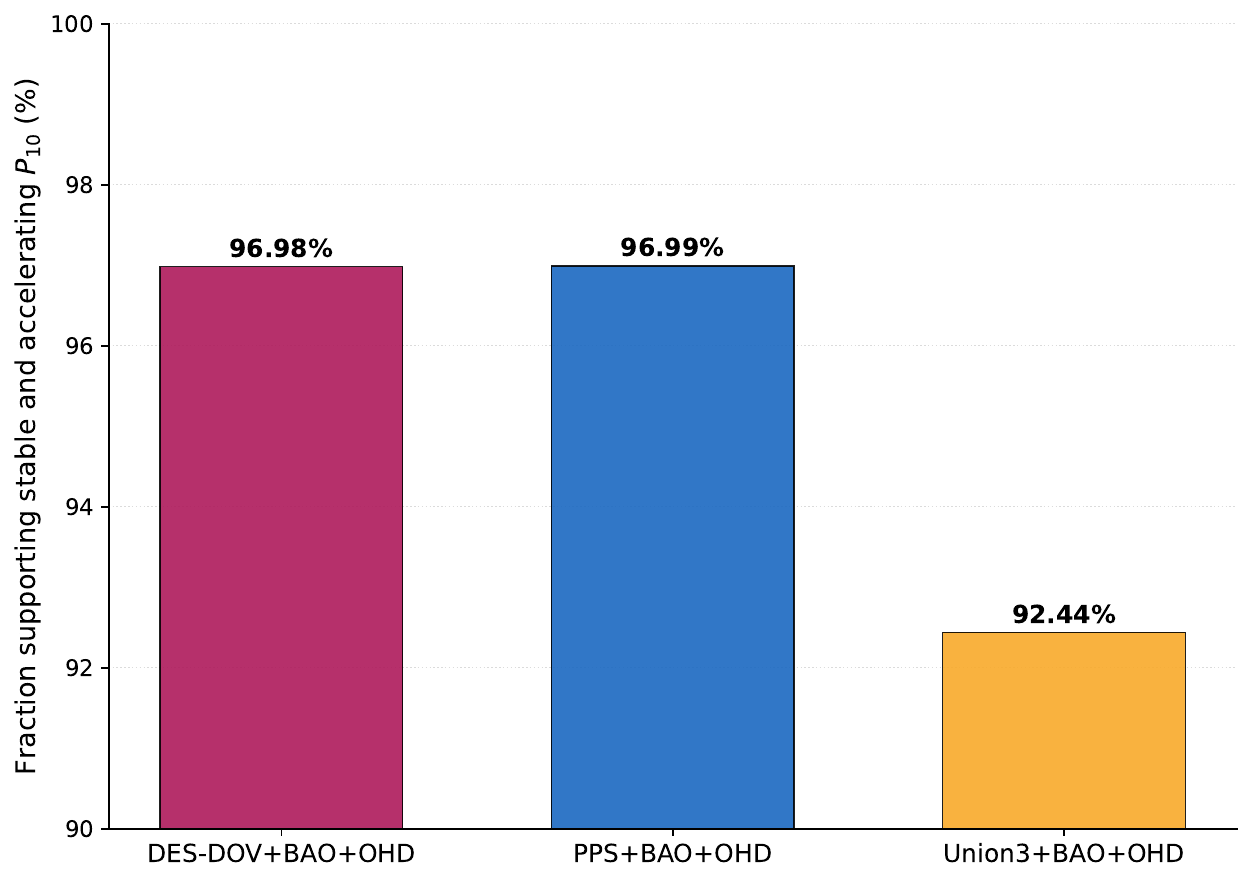}
        \caption{Posterior fractions satisfying
        stability and acceleration conditions.}
    \end{subfigure}
    \caption{Posterior-level dynamical assessment of
    the $P_{10}$ endpoint. The left panel shows the
    posterior distributions relative to the analytic
    stability boundary, while the right panel gives
    the fractions of retained MCMC samples corresponding
    to stable, accelerating solutions.}
    \label{fig:P10_pos}
\end{figure}

Figure~\ref{fig:P10_pos}(a) illustrates the posterior
distributions in the $(X,Y)$ plane, showing that most of
the allowed parameter space lies within the stability region
of the dark-energy-dominated attractor $P_{10}$.
The posterior fractions satisfying both stability and
acceleration conditions are approximately $96.98\%$ for
DES-DOV, $96.99\%$ for PPS, and $92.44\%$ for Union3
[Fig.~\ref{fig:P10_pos}(b)], with the Union3 posterior extending
relatively closer to the stability boundary.
These results highlight the dynamical significance of
including a separately conserved dark-energy component
within the proposed particle-creation framework.
In particular, the observationally favored solutions
consistently select a physically admissible, stable,
dark-energy-dominated accelerating attractor, while
particle creation modifies the effective DM dynamics
during cosmic evolution. The substantial posterior
support for $P_{10}$ further establishes the robustness
of this asymptotic configuration within the adopted model,
providing a consistent description of the distant cosmic
future characterized by quintessence-like accelerated
expansion with $H\rightarrow0$.

\subsection{Phase-space and density evolution}
\label{subsec:phase_density}

\begin{figure}[H]
    \centering

    \begin{subfigure}[t]{0.32\textwidth}
        \centering
        \includegraphics[width=\linewidth]{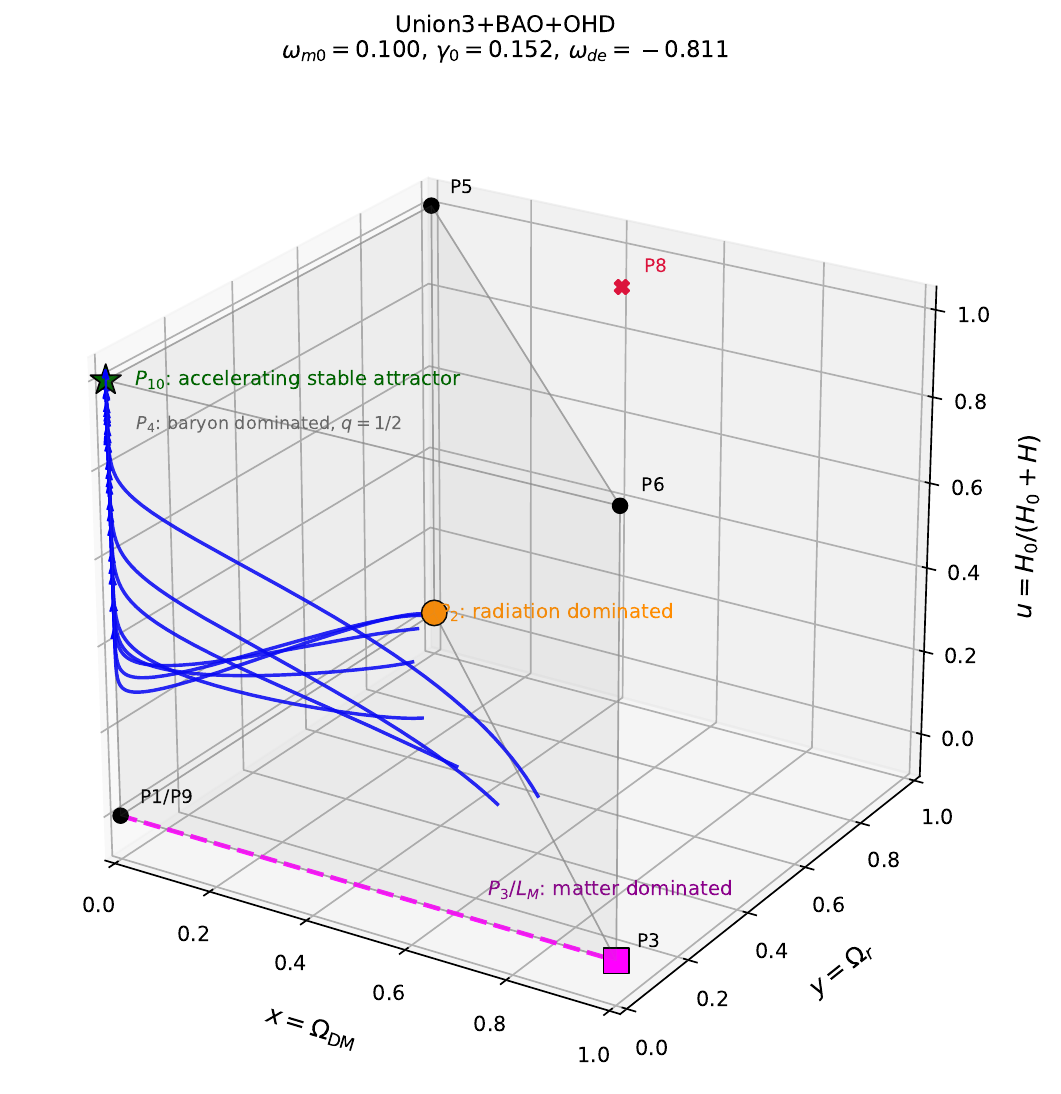}
        \caption{Union3+BAO+OHD}
    \end{subfigure}
    \hfill
    \begin{subfigure}[t]{0.32\textwidth}
        \centering
        \includegraphics[width=\linewidth]{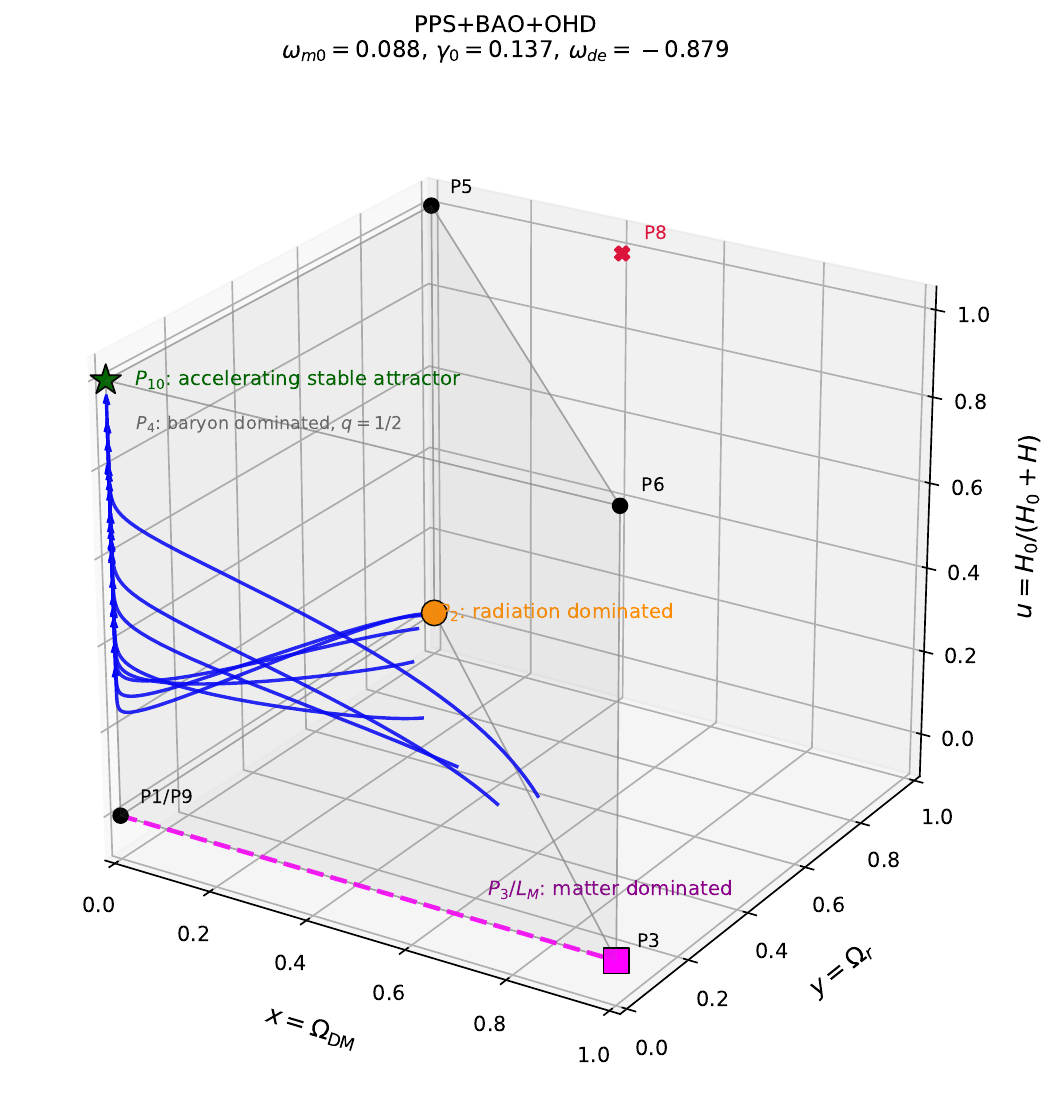}
        \caption{PPS+BAO+OHD}
    \end{subfigure}
    \hfill
    \begin{subfigure}[t]{0.32\textwidth}
        \centering
        \includegraphics[width=\linewidth]{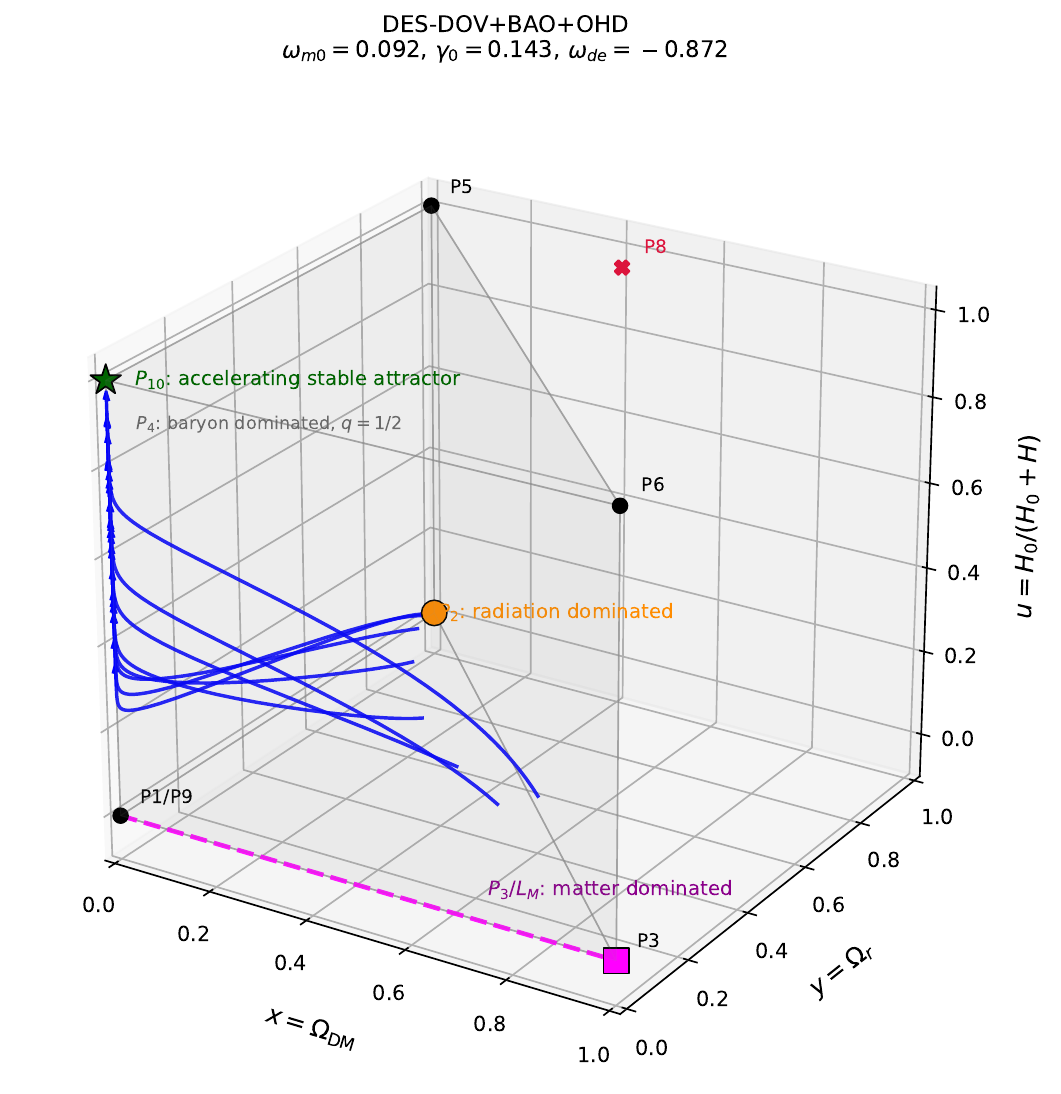}
        \caption{DES-DOV+BAO+OHD}
    \end{subfigure}

    \caption{
    Three-dimensional $(x,y,u)$ projections of
    representative trajectories obtained by numerically
    evolving the complete four-dimensional autonomous
    system $(x,y,b,u)$ for the three observational
    best-fit parameter sets. The trajectories illustrate
    the evolution from the matter-dominated region toward
    the late-time accelerating attractor, with increasing
    e-folding time $N=\ln a$.
    Radiation-, matter-, and dark-energy-dominated
    critical configurations are indicated.
    Critical points differing only in the suppressed
    baryonic coordinate $b$ overlap in the projected
    phase space.
    }
    \label{fig:phase_portraits_3datasets}
\end{figure}

The global dynamical behavior of the model is illustrated in
Fig.~\ref{fig:phase_portraits_3datasets}, where the complete
four-dimensional autonomous system $(x,y,b,u)$ is integrated
numerically and the resulting trajectories are projected onto
the three-dimensional $(x,y,u)$ subspace. Although the baryonic
degree of freedom is not explicitly displayed, it remains
dynamically active throughout the evolution. Three observational best-fit parameter sets exhibit
qualitatively similar phase-space trajectories, illustrating
the robustness of the underlying cosmological dynamics.
In the early high-$H$ regime, the normalized variable
$u$ approaches zero, and the effective dark-matter
equation of state satisfies

\begin{equation}
    \omega_{m,\mathrm{eff}}
    =2\Omega_{m0}u-4\Gamma_0u^2
    \longrightarrow 0.
\end{equation}

Consequently, the particle-creation-induced effective
pressure becomes negligible, and dark matter behaves
approximately as a pressureless fluid, similarly to
the conserved baryonic component. The trajectories
subsequently evolve through the neighborhood of the
matter-dominated critical line $L_M$, where the total
matter contribution dominates the total energy.
As the expansion proceeds and the Hubble parameter
decreases, the normalized variable $u$ increases,
enhancing the contribution of the effective dark-matter
pressure and modifying the background evolution.
The phase-space trajectories eventually converge
toward the dark-energy-dominated region associated
with the critical point $P_{10}$.

For the three observational best-fit parameter sets,
the dark-energy equation-of-state parameter lies
in the interval $ -1<\omega_{\mathrm{DE}}<-\frac{1}{3}$, 
corresponding to an accelerating dark-energy-dominated
asymptotic regime. The numerical convergence toward
$P_{10}$ is consistent with its attracting character
in the relevant parameter region. It is worth emphasizing that the three-dimensional representation suppresses the baryonic coordinate $b$.
Consequently, the critical-point pairs $(P_1,P_9)$
and $(P_4,P_{10})$, which share identical projected
coordinates, appear at the same locations in the
$(x,y,u)$ subspace, although they represent distinct
configurations in the complete four-dimensional
phase space.

\begin{figure}[t]
    \centering

    \begin{subfigure}[t]{0.485\linewidth}
        \centering
        \includegraphics[width=\linewidth]
        {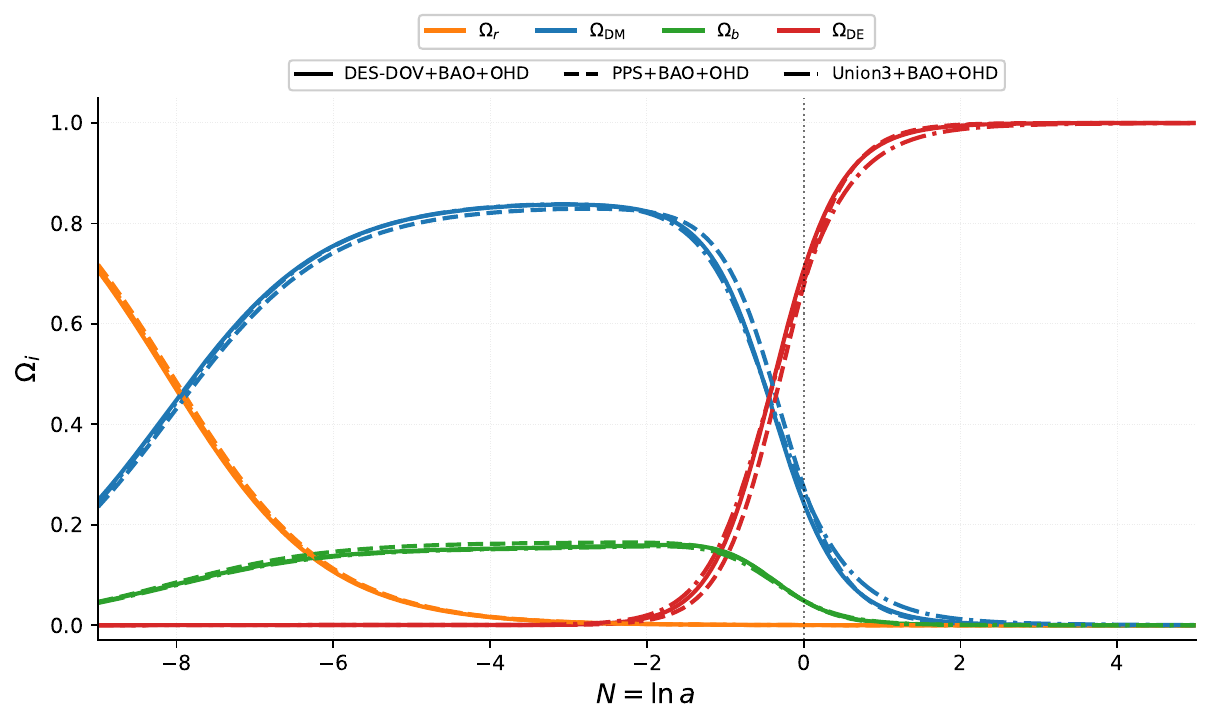}
        \caption{Evolution of the cosmic density fractions.}
    \end{subfigure}
    \hfill
    \begin{subfigure}[t]{0.485\linewidth}
        \centering
        \includegraphics[width=\linewidth]
        {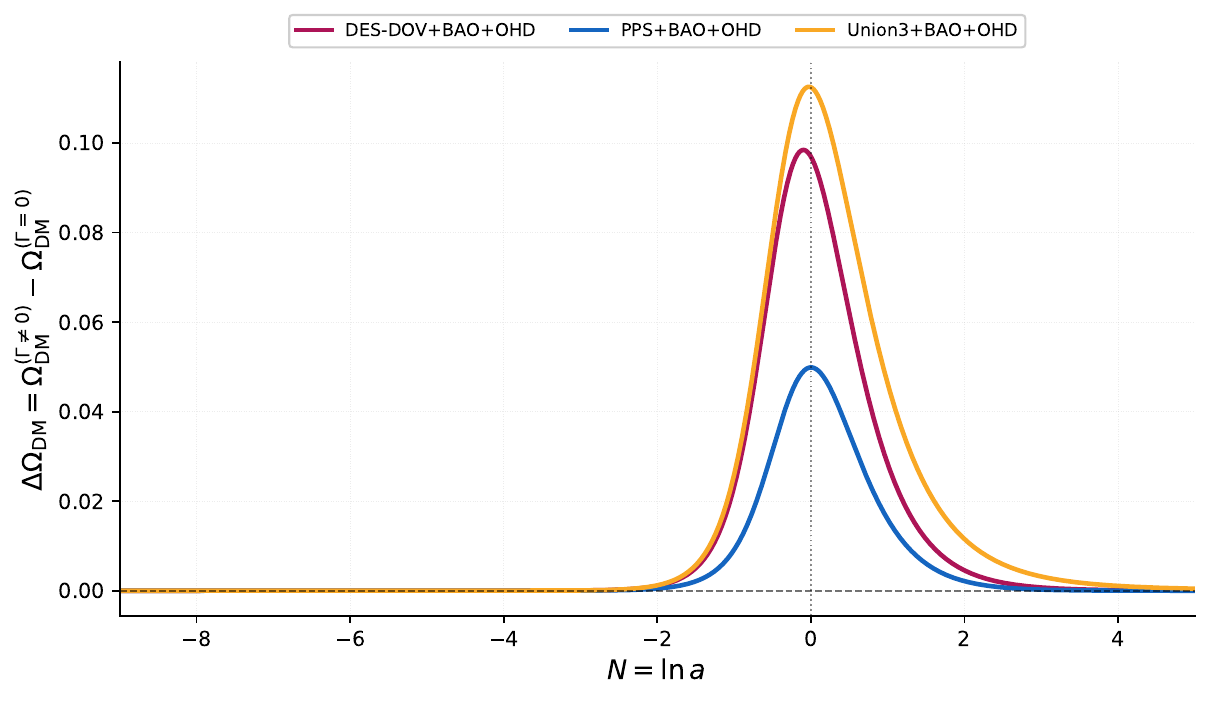}
        \caption{Creation-induced change in the DM density fraction.}
    \end{subfigure}

    \caption{
    Cosmological density evolution and the influence
    of the creation-related contribution for the three
    observational best-fit parameter sets.
    Panel (a) shows the radiation, dark-matter,
    baryonic, and dark-energy density fractions
    as functions of the e-folding variable $N=\ln a$.
    Panel (b) presents the difference
    $\Delta\Omega_{\mathrm{DM}}$ between the full model
    and the reference evolution with $\gamma_0=0$,
    using identical initial conditions and keeping
    the remaining best-fit parameters fixed.
    The differences peak near the present epoch
    and decrease toward both the early and
    late-time limits.
    }
    \label{fig:evo}
\end{figure}

The corresponding cosmological density evolution is
presented in Fig.~\ref{fig:evo}(a). Starting from
the observationally calibrated present-day configuration,
the dynamical equations are integrated backward and
forward in the e-folding variable $N=\ln a$.
The normalized density fractions satisfy

\begin{equation}
    \Omega_{\mathrm{DM}}=x,\qquad
    \Omega_r=y,\qquad
    \Omega_b=b,\qquad
    \Omega_{\mathrm{DE}}=1-x-y-b.
\end{equation}

The reconstructed histories reproduce the standard
cosmological sequence, beginning with radiation
domination at early times, followed by an extended
matter-dominated epoch and the eventual emergence
of dark-energy domination. During the intermediate matter-dominated regime,
the dark-matter fraction reaches approximately
$\Omega_{\mathrm{DM}}\simeq0.8$--$0.85$, while
the baryonic contribution remains around
$\Omega_b\simeq0.15$--$0.17$.
Both matter components subsequently decrease in
fractional abundance as dark energy becomes dominant.
In the asymptotic future, the solutions approach

\begin{equation}
    \Omega_{\mathrm{DE}}\longrightarrow1,
    \qquad
    \Omega_{\mathrm{DM}},
    \Omega_b,
    \Omega_r
    \longrightarrow0.
\end{equation}

The near coincidence of the density histories obtained
from Union3, PPS, and DES-DOV indicates that the
qualitative sequence of cosmological epochs is largely
insensitive to the choice of supernova compilation
within the best-fit parameter region considered. The influence of the creation-related contribution
on the dark-matter density evolution is further
quantified in Fig.~\ref{fig:evo}(b) through the
difference

\begin{equation}
    \Delta\Omega_{\mathrm{DM}}(N)
    =
    \Omega_{\mathrm{DM}}^{(\gamma_0\neq0)}(N)
    -
    \Omega_{\mathrm{DM}}^{(\gamma_0=0)}(N).
    \label{eq:delta_omega_dm}
\end{equation}

Here, the reference evolution is obtained by setting
$\gamma_0=0$, while the remaining best-fit parameters
and the calibrated initial conditions are kept fixed.
This construction isolates the modification associated
with the $\gamma_0$-dependent contribution to the
effective dark-matter equation of state. At early times, the difference is negligible because
the normalized variable $u$ is small, suppressing
the contribution proportional to $u^2$.
As the Universe evolves toward the present epoch,
the accumulated effect becomes appreciable,
producing a positive deviation in the dark-matter
density fraction relative to the reference evolution.
The three observational combinations exhibit
a similar qualitative behavior, with the difference
reaching its maximum near the present epoch.
The largest effect is obtained for Union3+BAO+OHD,
followed by DES-DOV+BAO+OHD and PPS+BAO+OHD.
The corresponding peak deviations are approximately
$0.11$, $0.10$, and $0.05$, respectively.
This ordering is consistent with the larger
best-fit creation parameters obtained for Union3
and DES-DOV, although the precise amplitude also
depends on the remaining background parameters
and their coupled dynamical evolution. Beyond the present epoch, the difference gradually
decreases and asymptotically approaches zero.
This behavior reflects the convergence of both
background solutions toward a dark-energy-dominated
state in which the fractional dark-matter abundance
vanishes. Importantly, the suppression of
$\Delta\Omega_{\mathrm{DM}}$ does not necessarily
indicate the disappearance of the normalized
particle-creation rate itself; rather, it reflects
the decreasing relative contribution of dark matter
to the total cosmic energy budget.

\section{Conclusions}
\label{sec:conclusions}

We have explored a late-time cosmological framework in which dark matter
is modeled as an effective open fluid incorporating both an intrinsic
pressure component and gravitationally induced particle creation.
The formulation is characterized by the compact variable $u$, which
maps the entire expanding cosmological history onto the bounded
interval $0\leq u\leq1$. Within this framework, the intrinsic
dark-matter equation of state depends linearly on $u$, while the
normalized particle-creation strength exhibits a quadratic dependence.
Consequently, both contributions become negligible in the early
high-$H$ regime, with the creation-induced term being suppressed
at a higher order in $u$. The model thus naturally recovers the
pressureless, non-creating dark-matter limit at early times, while
allowing the effects of particle creation to emerge progressively
during the subsequent cosmic evolution and become significant
at late times. 

The background evolution was formulated as a four-dimensional
autonomous dynamical system in terms of the variables
$(\Omega_{\rm DM},\Omega_r,\Omega_b,u)$. The corresponding
phase-space analysis reveals a rich dynamical structure,
comprising radiation-, matter-, and dark-energy-dominated
critical configurations, along with two nontrivial
dark-matter branches. A detailed Jacobian analysis
further identifies a continuous line of matter-dominated
critical points, arising from the dynamical degeneracy
between baryons and dark matter in the early-time limit
$\omega_{m,\rm eff}\rightarrow0$. Furthermore, the
late-time stability analysis establishes that the
existence of an accelerating critical point alone
does not guarantee a physically viable cosmic future.
Rather, the identification of a physically admissible
late-time attractor requires the simultaneous fulfillment
of the accelerated-expansion condition, dynamical
stability criteria, and the constraints imposed by
the compact physical phase space.

 The distinction between mathematical existence and physical
admissibility is crucial for identifying the late-time attractor.
For $\gamma_0>0$, the nontrivial dark-matter point $P_7$ lies
outside the physical phase space ($u_7<0$), while $P_8$,
although potentially stable and admissible in a restricted
parameter region, satisfies $u_8>1$ for all three observational
best-fit sets. Thus, neither branch represents a physically
viable late-time solution despite its formal de Sitter-like
behavior. Among the remaining accelerating configurations,
$P_6$ and $P_9$ exhibit saddle behavior in the observationally
favored parameter region. In contrast, the dark-energy-dominated
point $P_{10}=(0,0,0,1)$ is physically admissible, accelerating,
and linearly stable for all three best-fit parameter sets,
thereby emerging as the common late-time attractor of the model.

 The physical interpretation of the late-time attractor $P_{10}$
is particularly significant. Since $u\rightarrow1$ corresponds
to $H\rightarrow0$, this equilibrium does not represent a
finite-$H$ de Sitter state. Instead, the observationally
favored range $-1<\omega_{\rm de}<-1/3$ leads to an
asymptotically accelerating, quintessence-like expansion
with a decreasing Hubble rate, identifying $P_{10}$ as
a dark-energy-dominated power-law endpoint. The observational
constraints from PPS+BAO+OHD, Union3+BAO+OHD, and
DES-DOV+BAO+OHD show broad consistency, yielding similar
matter densities and background expansion histories, with
moderate differences primarily in $h$ and $\omega_{\rm de}$.
All three combinations favor a mildly negative present-day
effective dark-matter equation of state, while the intrinsic
parameter $\omega_{m0}=\omega_{m,\rm eff}^{0}+\gamma_0$
remains positive. This reflects the interplay between
positive intrinsic dark-matter pressure and negative
creation pressure, producing a pronounced anticorrelation
between $\omega_{m,\rm eff}^{0}$ and $\gamma_0$ in the
posterior distributions.

The reconstructed background evolution exhibits remarkable
consistency across the three supernova compilations, with
a nearly identical deceleration-to-acceleration transition
at $z_{\rm tr}\simeq0.71$--$0.73$, coinciding with the
crossing of the acceleration threshold by the total
equation of state. At high redshifts, the intrinsic
dark-matter equation of state approaches zero, while
the normalized particle-creation strength becomes
strongly suppressed, preserving the standard early-time
matter-dominated behavior. Beyond the best-fit analysis,
the posterior-level dynamical assessment reveals that
approximately $96.98\%$, $96.99\%$, and $92.44\%$ of the
retained DES-DOV, PPS, and Union3 posterior samples,
respectively, satisfy the combined stability and
acceleration conditions for $P_{10}$. The slightly
lower fraction for Union3 reflects its posterior
distribution extending closer to the analytical
stability boundary rather than a qualitatively
different asymptotic evolution. These findings
demonstrate that the physically admissible,
accelerating late-time attractor is supported
throughout the majority of the observationally
allowed parameter space, reinforcing the
dynamical robustness of the proposed framework. 

 The global phase-space trajectories and reconstructed density
histories consistently reproduce the standard cosmological
sequence, evolving from radiation domination through an
extended matter-dominated epoch toward a late-time accelerating,
dark-energy-dominated regime. The creation-related modification
of the dark-matter fraction is negligible in the early high-$H$
limit, becomes appreciable during the late-time transition,
and reaches its maximum near the present epoch. Its subsequent
suppression reflects the diminishing fractional abundance
of dark matter rather than the disappearance of the normalized
particle-creation strength. Collectively, these findings establish
a consistent connection between the effective open-fluid
description, global phase-space dynamics, and observational
constraints. While the model preserves the conventional
early-time cosmological evolution and accommodates nontrivial
particle-creation effects at late times, the combined stability
and physical phase-space conditions identify $P_{10}$ as the
common viable accelerating attractor for all three observational
best-fit sets, excluding the formally de Sitter-like
dark-matter branches from the physically realized
asymptotic evolution.

The dynamical significance of the dark-energy component is
further established by the observationally calibrated results.
Fig.~\ref{fig:critical_point_summary} identifies the dark-energy-dominated point $P_{10}$
as the physically admissible stable accelerating attractor,
while Fig.~\ref{fig:P10_pos} demonstrates its robustness across the posterior
distributions, with support exceeding $92\%$ for all three
datasets. The reconstructed density evolution in Fig.~\ref{fig:evo}
further confirms the transition from radiation and matter
domination toward asymptotic dark-energy domination.
Consequently, the framework characterizes not only the
late-time accelerating expansion but also the distant
cosmic future, approaching a quintessence-like accelerating
state with a decreasing Hubble parameter.

The present investigation is primarily devoted to the background
cosmological dynamics, phase-space structure, and observational
viability of the proposed particle-creation framework. Although
the analysis demonstrates a consistent cosmological evolution
at the homogeneous level, a complete treatment of linear
cosmological perturbations remains an important extension of
the model. In particular, the expansion-dependent effective
dark-matter pressure may introduce nonadiabatic perturbations
and modify the growth of cosmic structures. A systematic
perturbative investigation would therefore require a
consistent covariant formulation of the particle-creation
mechanism, including the associated pressure perturbations,
entropy production, and momentum dynamics. Such a framework
would enable a detailed assessment of the effective sound
speed, scalar perturbative stability, and matter-growth
history, as well as comparisons with large-scale-structure
observables such as $f\sigma_8$, weak gravitational lensing,
and CMB anisotropies
\cite{Sagredo:2018ahx,Kazantzidis:2018rnb,Linder:2003dr}.
These investigations are essential for determining whether
the background-level viability established here extends
to perturbative stability and compatibility with
structure-formation constraints.

\appendix
\section{Higher-Order Extension of the Compact Dark-Matter Model}
\label{app:higher_order}

The linear--quadratic parametrization adopted in the main
analysis represents a minimal realization of the effective
dark-matter sector in terms of the compact variable
\begin{equation}
u=\frac{H_0}{H_0+H}.
\end{equation}
A systematic higher-order generalization can be introduced
by expanding the intrinsic dark-matter equation of state
and the normalized particle-creation rate as
\begin{align}
\omega_m(u)
&=2\omega_{m0}u+\sum_{n=2}^{N}\alpha_n u^n,
\\
\frac{\Gamma}{3H}
&=4\gamma_0u^2+\sum_{n=3}^{M}\beta_n u^n.
\end{align}

The corresponding effective dark-matter equation of state
becomes
\begin{equation}
\omega_{m,\mathrm{eff}}(u)
=
2\omega_{m0}u-4\gamma_0u^2
+\sum_{n=2}^{N}\alpha_nu^n
-\sum_{n=3}^{M}\beta_nu^n.
\label{eq:weff_higher}
\end{equation}
Since all additional contributions vanish as
$u\rightarrow0$, the pressureless and non-creating
high-$H$ limit remains unchanged.

Retaining terms up to cubic order gives
\begin{equation}
\omega_{m,\mathrm{eff}}(u)
=
2\omega_{m0}u
+(\alpha_2-4\gamma_0)u^2
+(\alpha_3-\beta_3)u^3,
\end{equation}
which, at the present epoch ($u_0=1/2$), reduces to
\begin{equation}
\omega_{m,\mathrm{eff}}^0
=
\omega_{m0}-\gamma_0
+\frac{\alpha_2}{4}
+\frac{\alpha_3-\beta_3}{8}.
\end{equation}

Although this extension provides a more flexible
description of the dark-matter sector, introducing
additional coefficients into the MCMC analysis
would enlarge the parameter space and potentially
generate significant degeneracies. In particular,
the background expansion depends on the combined
effective equation of state
$\omega_{m,\mathrm{eff}}(u)$, making it difficult
to distinguish the individual contributions of
the intrinsic-pressure and particle-creation
coefficients using background observables alone.
For example, the cubic coefficients $\alpha_3$
and $\beta_3$ enter only through their difference,
leading to an exact degeneracy at the background
level unless additional physical information
is supplied. An improvement in the minimum $\chi^2$ alone does not justify additional model complexity, since higher-order parameters increase fitting flexibility. Model-selection criteria such as AIC and BIC penalize extra parameters, requiring a sufficiently large improvement in the best-fit likelihood to favor the extended model. From a dynamical perspective, polynomial extensions
in the existing variable $u$ do not introduce
additional autonomous degrees of freedom when
their coefficients are constants. The phase-space
dimension therefore remains four, described by
$(\Omega_{\mathrm{DM}},\Omega_r,\Omega_b,u)$.
Nevertheless, the resulting fixed-point equations
become higher-order polynomials, potentially
introducing additional critical-point branches
and more restrictive physicality and stability
conditions. Extensions involving new dynamical
fields or independently evolving creation variables
would instead enlarge the autonomous system and
require a correspondingly extended stability analysis.
These considerations motivate retaining the
linear intrinsic equation of state and quadratic
particle-creation contribution as the minimal
working parametrization. This choice preserves
the conventional early-time matter behavior
while allowing nontrivial modifications to
the late-time expansion history without
introducing poorly constrained higher-order
coefficients. A systematic investigation of
the extended framework, incorporating additional
background and perturbation-level observables,
including growth-rate and large-scale-structure
measurements, is left for future work.

\paragraph*{\textit{Data Availability Statement}} All original datasets used in this work are publicly accessible and have been appropriately cited in the bibliography.

\end{document}